\documentclass[useAMS,usenatbib]{mn2e}

\usepackage[dvips]{graphicx}   
\usepackage{mathrsfs}

\usepackage{natbib}
\def\reff@jnl#1{{\rm#1\/}}
\def\aj{\reff@jnl{AJ}}                  
\def\araa{\reff@jnl{ARA\&A}}            
\def\apj{\reff@jnl{ApJ}}                
\def\apjl{\reff@jnl{ApJ}}               
\def\apjs{\reff@jnl{ApJS}}              
\def\apss{\reff@jnl{Ap\&SS}}            
\def\aap{\reff@jnl{A\&A}}               
\def\aapr{\reff@jnl{A\&A~Rev.}}         
\def\aaps{\reff@jnl{A\&AS}}             
\def\mnras{\reff@jnl{MNRAS}}            
\def\prd{\reff@jnl{Phys.Rev.D}}         
\def\prl{\reff@jnl{Phys.Rev.Lett}}      
\def\pasp{\reff@jnl{PASP}}              
\def\pasj{\reff@jnl{PASJ}}              
\def\nat{\reff@jnl{Nature}}             

\newcommand{\bd}{\begin{displaymath}}
\newcommand{\ed}{\end{displaymath}}
\newcommand{\be}{\begin{equation}}
\newcommand{\ee}{\end{equation}}
\newcommand{\beaa}{\begin{eqnarray*}}
\newcommand{\eeaa}{\end{eqnarray*}}
\newcommand{\bea}{\begin{eqnarray}}
\newcommand{\eea}{\end{eqnarray}}

\def\dataVec{\mathbfit{d}}
\def\data{d}
\def\responseSet{\mathbfss{f}}

\def\regSet{\mathbfss{g}}

\def\srVec{\mathbfit{s}}

\def\srMPVec{\mathbfit{s}_{\mathrm{MP}}}
\def\srMLVec{\mathbfit{s}_{\mathrm{ML}}}

\def\noiseVec{\mathbfit{n}}

\def\imCM{\mathbfss{C}_{\mathrm{D}}}

\def\priorCM{\mathbfss{S}}
\def\trueCM{\mathbfss{S}_*}

\def\hessD{\mathbfss{B}}
\def\hessS{\mathbfss{C}}
\def\hessM{\mathbfss{A}}

\def\IdenM{\mathbfss{I}}

\title%
[Bayesian source inversion in gravitational lensing]%
{A Bayesian analysis of regularised source inversions in gravitational lensing}
\label{firstpage}

\author%
[S. H. Suyu et al.]%
{S. H. Suyu$^{1,2}$\thanks{E-mail:suyu@its.caltech.edu},
  P. J. Marshall$^{2}$,
  M. P. Hobson$^{3}$, and
  R. D. Blandford$^{1,2}$ \\
$^{1}$Theoretical Astrophysics, 103-33, California Institute of Technology,
  Pasadena, CA, 91125, USA \\ 
$^{2}$KIPAC, Stanford University, 2575 Sand Hill Road, Menlo Park, CA 94025,
  USA \\ 
$^{3}$Astrophysics Group, Cavendish Laboratory, Madingley Road, Cambridge CB3
  0HE, UK} 
\date{Accepted ---; received ---; in original form \today}

\pagerange{\pageref{firstpage}--\pageref{lastpage}}

\pubyear{2005}

\begin{document}

\maketitle


\begin{abstract}

Strong gravitational lens systems with extended sources are of special
interest because they provide additional constraints on the models of the
lens systems.  To use a gravitational lens system for measuring the Hubble
constant, one would need to determine the lens potential and
the source intensity distribution simultaneously. A linear inversion method to
reconstruct a 
pixellated source brightness distribution of a given lens potential model was introduced
by Warren \& Dye.  In the inversion process, a regularisation on the source
intensity is often needed to ensure a successful inversion with a faithful
resulting source.  
In this paper, we use Bayesian analysis to determine the optimal
regularisation constant (strength of regularisation) of a given form of
regularisation and to
objectively choose the optimal form of regularisation given a selection of
regularisations.  We consider and compare quantitatively three different forms
of regularisation previously described in the literature for source inversions
in 
gravitational lensing: zeroth-order, gradient and curvature.  We use simulated
data with the exact lens potential to demonstrate the method.  We find that
the preferred form of regularisation depends on the nature of the source
distribution.  

\end{abstract}



\section{Introduction} 

The use of strong gravitational lens systems to measure cosmological
parameters and to probe matter (including dark matter) is well known
\citep*[e.g.\ ][]{R64,KSW04}.  Lens systems with extended source brightness distributions
are particularly useful since they provide additional constraints for the lens
modelling due to surface brightness conservation.  In such a system, one would
need to fit simultaneously the source intensity distribution and the lens
potential model (or, equivalently the lens mass distribution) to the
observational data.  The use of a pixellated source brightness distribution has the
advantage over a parametric source brightness distribution in that the source model is
not restricted to a particular parameter space.  \citet{WD03} introduced a
linear inversion method to obtain the best-fitting pixellated source
distribution given a lens model and the observational data.  Several groups of
people \citep*[e.g.\ ][]{W96, TK04, DW05, K05, BL05} have used pixellated source
distributions, and some \citep{K05, SB05} even used a pixellated potential model
for the lens.

The method of source inversion described in \citet{WD03} requires the source
distribution to be ``regularised'' (i.e., smoothness conditions on the
inverted source intensities to be imposed) for reasonable source
resolutions.\footnote{The source pixel sizes are fixed and are roughly a factor
of the average magnification smaller than the image pixel sizes.  
In this case, regularisation is needed because the number of source
pixels is comparable to the number of data pixels.  On the other hand,
if the number of source pixels is much fewer than the effective number
of data pixels (taking into account of the signal-to-noise ratio), the
data alone could be sufficient to constrain the pixellated source
intensity values and regularisation would play little role.  This is
equivalent to imposing a uniform prior on the source intensity
distribution (a prior on the source is a form of regularisation), a point to
which we will return later in this article.}  For fixed
pixel sizes, there are various forms of regularisation to use and the
differences among them have not been addressed in detail.  In addition,
associated with a given form of regularisation is a regularisation constant
(signifying the strength of the regularisation), and the way to set this
constant has been unclear.  These two long-standing problems were noted in
\citet{KSW04}.  Our goal in this paper is to use Bayesian analysis
to address the above two issues by quantitatively comparing different values
of the regularisation constant and the forms of regularisation.  

\citet{BL05} also followed a Bayesian approach for pixellated source
inversions.  The main difference between \citet{BL05} and this paper is the
prior on the source intensity distribution.  Furthermore, this paper quantitatively compares the various
forms of regularisation by evaluating the so-called ``evidence'' for each of
the forms of regularisation in the Bayesian framework; \citet{BL05} mentioned
the concept of model comparison but did not apply it.

\citet{DW05} use adaptive source grids to avoid the use of explicit regularisation (i.e., uniform priors are imposed since adapting the grids is an implicit form of regularisation);
however, the Bayesian formalism would still be useful to set the optimal
scales of the adaptive pixel sizes objectively.  Furthermore, regularised source
inversions (as opposed to unregularised -- see footnote 1) 
permit the use of smaller pixel sizes
to obtain fine structures.

The outline of the paper is as follows.  In Section \ref{sect:BayesInf}, we introduce the theory
of Bayesian inference, describing how to fit a model to a given set of data
and how to rank the various models.  In Section \ref{sect:appgl}, we apply the Bayesian
analysis to source inversions in strong gravitational lensing and show a way to
 rank the different forms of regularisations quantitatively.


\section{Bayesian Inference}
\label{sect:BayesInf}

We follow \citet{M92} for the theory of Bayesian analysis, but use different
notations that are more convenient for the application to gravitational
lensing in Section \ref{sect:appgl}.  

In Bayesian analysis, there are two levels of inference for data modelling. 
In the first level of inference, we choose a model and fit it to the data. 
This means characterising the probability distribution for the 
parameters of the model given the
data.  In the second level of inference, we want to rank the
models quantitatively in the light of the data.  By asking for the relative
probabilities of 
models given the data, Bayesian analysis incorporates Occam's razor
(which states that overly complex models should not be preferred over simpler
models unless the data support them) in this second level of inference.  
The appearance of Occam's razor
will be evident at the end of Section \ref{sect:BayesInf:ModelComp:findLambda}.  In the following subsections, we
will describe the two levels of inference in detail.


\subsection{Model fitting}
\label{sect:BayesInf:ModelFit}

Let $\dataVec$ be a vector of data points $\data_j$, where $j=1,\ldots,N_{\rm d}$ and $N_{\rm d}$
is the total number of data points.  Let $s_i$ be the model parameters that we
want to infer given the data, where $i=1,\ldots,N_{\rm s}$ and $N_{\rm s}$ is the number of
parameters.  Let $\responseSet$ represent the response function that relates
the model parameters to the measured data. (In the application of source
reconstruction in gravitational lensing in Section \ref{sect:appgl},
$\responseSet$ encodes information on the lens potential, which is fixed in
each iteration of source reconstruction.) For simplicity, consider
$\responseSet$ to be a constant linear transformation matrix of dimensions $N_{\rm d}$-by-$N_{\rm s}$ such that \be
\label{eq:dj}
\dataVec = \responseSet \srVec + \noiseVec
\ee
where $\noiseVec$ is the noise in the data characterised by the covariance matrix
$\mathbfss{C}_{\mathrm{D}}$ (here and below, subscript D indicates ``data'').

Modelling the noise as Gaussian,\footnote{The Gaussian assumption is usually 
applicable to optical CCD data which
have noise at each pixel characterised by dispersion $\sigma_j$, the
square root of the corresponding diagonal entry of the covariance matrix.  In
general, there is correlation between adjacent pixels due to charge transfer 
(bleeding) and the drizzling process, which is characterised by the
off-diagonal terms in the covariance matrix.}
the probability of
the data given the model parameters 
$\mathbfit{s}$ is 
\be
\label{eq:likelihood}
P(\dataVec | \mathbfit{s}, \responseSet) = \frac {\exp(-E_{\mathrm{D}}(\dataVec | \mathbfit{s},\responseSet))}{Z_{\mathrm{D}}},
\ee
where 
\bea
\label{eq:ED}
E_{\mathrm{D}}(\dataVec | \mathbfit{s},\responseSet) &=& \frac{1}{2} \left(\responseSet \srVec - \dataVec \right)^{\mathrm{T}} \mathbfss{C}_{\mathrm{D}}^{-1} \left(\responseSet \srVec - \dataVec \right)
\nonumber \\ & = &
\frac{1}{2}\chi^2
\eea
and $Z_{\mathrm{D}} = (2\pi)^{N_{\rm d}/2} (\det \mathbfss{C}_{\mathrm{D}})^{1/2}$
is the normalisation for the probability. The probability $P(\dataVec |
\mathbfit{s}, \responseSet)$ is called the \textit{likelihood}, and
$E_{\mathrm{D}}(\dataVec | \mathbfit{s},\responseSet)$ is half the standard
value of $\chi^2$.  In many cases, the problem of finding the most likely
solution $\mathbfit{s}_{\mathrm{ML}}$ that minimizes $E_{\mathrm{D}}$ is
ill-posed.  This indicates the need to set a \textit{prior} $P(\mathbfit{s}|\regSet, \lambda)$ on the
parameters $\mathbfit{s}$.  The prior can be thought of as ``regularising''
the parameters $\mathbfit{s}$ to make the prediction $
\responseSet\srVec$ smooth.  We can express the prior in the following form

\be
\label{eq:prior}
P(\mathbfit{s}|\regSet, \lambda) = \frac {\exp (-\lambda E_{\mathrm{S}}(\mathbfit{s}|\regSet))}{Z_{\mathrm{S}}(\lambda)},
\ee
where $\lambda$, the so-called regularisation constant, is the strength of
regularisation and $Z_{\mathrm{S}}(\lambda)=\int \mathrm{d}^{N_{\rm s}}\mathbfit{s}
\exp(-\lambda E_{\mathrm{S}})$ is the normalisation of the prior probability
distribution.  The function $E_{\mathrm{S}}$ is often called the regularising
function.  We focus on commonly used quadratic forms of the regularising
function, and defer the discussion of other priors to Section \ref{sect:BayesInf:ModelComp:rankModel}. 
As we will see in Section \ref{sect:BayesInf:ModelComp:findLambda}, Bayesian analysis
allows us to infer quantitatively the value of $\lambda$ from the data in the
second level of inference.  

Bayes' rule tells us that the \textit{posterior probability} of the parameters
$\mathbfit{s}$ given the data, response function and prior is 
\be 
\label{eq:posterior}
P(\mathbfit{s}|\dataVec,\lambda,\responseSet,\regSet) = \frac{P(\dataVec | \mathbfit{s}, \responseSet) P(\mathbfit{s}|\regSet, \lambda)}{P(\dataVec|\lambda,\responseSet, \regSet)},
\ee
where $P(\dataVec|\lambda,\responseSet, \regSet)$ is the normalisation that is
called the \textit{evidence} for the model
$\{\lambda,\responseSet,\regSet\}$.  Since both the likelihood and
prior are either approximated or set as Gaussians, the posterior probability
distribution is also a Gaussian.  The evidence is irrelevant in the first
level of inference where we maximize the posterior (equation
(\ref{eq:posterior})) of parameters $\mathbfit{s}$ to obtain the most 
probable parameters
$\mathbfit{s}_{\mathrm{MP}}$.  However, the evidence is important in the
second level of inference for model comparisons.  Examples of using the
evidence in astronomical context are \citet*{HBL02} and \citet{Metal02}.

To simplify the notation, let us define 
\be 
\label{eq:M}
M(\mathbfit{s}) = E_{\mathrm{D}}(\mathbfit{s}) + \lambda E_{\mathrm{S}}(\mathbfit{s}).
\ee
With the above definition, we can write the posterior as 
\be
\label{eq:posterior2}
P(\mathbfit{s}|\dataVec,\lambda,\responseSet,\regSet) = \frac{\exp(-M(\mathbfit{s}))}{Z_{\mathrm{M}}(\lambda)},
\ee
where $Z_{\mathrm{M}}(\lambda) = \int \mathrm{d}^{N_{\rm s}} \mathbfit{s} \exp(-M(\mathbfit{s}))$ is the normalisation.


\subsection*{The most likely versus the most probable solution}

By definition, the most likely solution $\mathbfit{s}_{\mathrm{ML}}$ maximizes
the likelihood, whereas the most probable solution
$\mathbfit{s}_{\mathrm{MP}}$ maximizes the posterior.  In other words,
$\mathbfit{s}_{\mathrm{ML}}$ minimizes $E_{\mathrm{D}}$ in equation
(\ref{eq:ED}) ($\mathbf{\nabla}
E_{\mathrm{D}}(\mathbfit{s}_{\mathrm{ML}})=\mathbf{0}$, where $\mathbf{\nabla}
\equiv \frac{\mathrm{\partial}}{\mathrm{\partial} \mathbfit{s}}$) and
$\mathbfit{s}_{\mathrm{MP}}$ minimizes $M$ in equation (\ref{eq:M})
($\mathbf{\nabla} M(\mathbfit{s}_{\mathrm{MP}})=\mathbf{0}$).

Using the definition of the most likely solution, it is not difficult to
verify that it is
\be 
\label{eq:sML}
\mathbfit{s}_{\mathrm{ML}} = \mathbfss{F}^{-1} \mathbfit{D},
\ee
where
\be
\label{eq:F}
\mathbfss{F} = \responseSet^{\mathrm{T}} \mathbfss{C}_{\mathrm{D}}^{-1} \responseSet
\ee
and
\be
\label{eq:D}
\mathbfit{D}= \responseSet^{\mathrm{T}} \mathbfss{C}_{\mathrm{D}}^{-1} \dataVec
\ee
The matrix $\mathbfss{F}$ is square with dimensions $N_{\rm s} \times N_{\rm s}$ and the
vector $\mathbfit{D}$ has dimensions $N_{\rm s}$.

The most probable solution $\mathbfit{s}_{\mathrm{MP}}$ can in fact be
obtained from the most likely solution $\mathbfit{s}_{\mathrm{ML}}$.  If the
regularising function $E_{\mathrm{S}}$ is a quadratic functional that obtains
its minimum at $\mathbfit{s}_{\mathrm{reg}}$ (i.e., $\mathbf{\nabla}
E_{\mathrm{S}}(\mathbfit{s}_{\mathrm{reg}})=\mathbf{0}$), then we can Taylor
expand $E_{\mathrm{D}}$ and $E_{\mathrm{S}}$ to
\be
\label{eq:EDtaylor}
E_{\mathrm{D}}(\mathbfit{s}) = E_{\mathrm{D}}(\mathbfit{s}_{\mathrm{ML}}) + \frac{1}{2} (\mathbfit{s}-\mathbfit{s}_{\mathrm{ML}})^{\mathrm{T}} \mathbfss{B} (\mathbfit{s}-\mathbfit{s}_{\mathrm{ML}})
\ee
and
\be
\label{eq:EStaylor}
E_{\mathrm{S}}(\mathbfit{s}) = E_{\mathrm{S}}(\mathbfit{s}_{\mathrm{reg}}) + \frac{1}{2} (\mathbfit{s}-\mathbfit{s}_{\mathrm{reg}})^{\mathrm{T}} \mathbfss{C} (\mathbfit{s}-\mathbfit{s}_{\mathrm{reg}}),
\ee
where $\mathbfss{B}$ and $\mathbfss{C}$ are Hessians of $E_{\mathrm{D}}$ and
$E_{\mathrm{S}}$, respectively: $\mathbfss{B}=\nabla \nabla
E_{\mathrm{D}}(\mathbfit{s})$ and $\mathbfss{C}=\nabla \nabla
E_{\mathrm{S}}(\mathbfit{s})$.  Equations (\ref{eq:EDtaylor}) and
(\ref{eq:EStaylor}) are exact for quadratic forms of $E_{\mathrm{D}}$ and
$E_{\mathrm{S}}$ with the Hessians $\mathbfss{B}$ and $\mathbfss{C}$ as
constant matrices.  For the form of $E_{\mathrm{D}}$ in equation
(\ref{eq:ED}), $\mathbfss{B}$ is equal to $\mathbfss{F}$ that is given by
equation (\ref{eq:F}).  We define $\mathbfss{A}$ as the Hessian of $M$, i.e.
 $\mathbfss{A}=\nabla
\nabla M(\mathbfit{s})$, and by equation (\ref{eq:M}), $\mathbfss{A}=
\mathbfss{B}+\lambda\mathbfss{C}$.  Using equations (\ref{eq:M}),
(\ref{eq:EDtaylor}), and (\ref{eq:EStaylor}) in $\mathbf{\nabla}
M(\mathbfit{s}_{\mathrm{MP}})=\mathbf{0}$, we can get the most probable
solution (that maximizes the posterior) as $\mathbfit{s}_{\mathrm{MP}} =
\mathbfss{A}^{-1} (\mathbfss{B} \mathbfit{s}_{\mathrm{ML}} + \lambda
\mathbfss{C} \mathbfit{s}_{\mathrm{reg}})$. The simplest forms of the prior,
especially the ones we will use for the gravitational lensing inversion in Section \ref{sect:appgl}, have
$\mathbfit{s}_{\mathrm{reg}}=\mathbf{0}$. 
In the case where $\mathbfit{s}$ correspond to
pixel intensity values, $\mathbfit{s}_{\mathrm{reg}}=\mathbf{0}$
implies a prior preference towards a blank image. The noise suppression effect
of the regularisation follows from this supplied bias.  
Focusing on such forms of prior, the most probable solution becomes
\be 
\label{eq:sMP}
\mathbfit{s}_{\mathrm{MP}} = \mathbfss{A}^{-1} \mathbfss{B} \mathbfit{s}_{\mathrm{ML}}.
\ee
This result agrees with equation (12) in \citet{WD03}.  In fact, equation
(\ref{eq:sMP}) is always valid when the regularising function can be written
as $E_{\mathrm{S}}(\mathbfit{s}) = \frac{1}{2} \mathbfit{s}^{\mathrm{T}}
\mathbfss{C} \mathbfit{s}$.

Equation (\ref{eq:sMP}) indicates a one-time calculation of
$\mathbfit{s}_{\mathrm{ML}}$ via equation (\ref{eq:sML}) that permits the
computation of the most probable solution $\mathbfit{s}_{\mathrm{MP}}$ by
finding the optimal regularisation constant of a given form of regularisation.
The parameters $\mathbfit{s}_{\mathrm{MP}}$ in equation
(\ref{eq:sMP}) depend on the regularisation constant $\lambda$ since the
Hessian $\mathbfss{A}$ depends on $\lambda$.  Bayesian analysis provides a
method for setting the value of $\lambda$, as described in the next subsection.


\subsection{Model comparison}
\label{sect:BayesInf:ModelComp}

In the previous section, we found that for a given set of data $\dataVec$ and
a model (response function $\responseSet$ and regularisation $\regSet$ with
regularisation constant $\lambda$), we could calculate the most probable
solution $\mathbfit{s}_{\mathrm{MP}}$ for the particular $\lambda$.  In this
section, we consider two main points: (i) how to set the regularisation
constant $\lambda$ for a given form of regularisation $\regSet$ and (ii) how
to rank the different models $\responseSet$ and $\regSet$.


\subsubsection{Finding $\lambda$}
\label{sect:BayesInf:ModelComp:findLambda}

To find the optimal regularisation constant $\lambda$, we want to maximize 
\be
\label{eq:lamProb}
P(\lambda | \dataVec, \responseSet, \regSet) = \frac{P(\dataVec|\lambda, \responseSet, \regSet) P(\lambda)}{P(\dataVec|\responseSet, \regSet)},
\ee
using Bayes' rule.  Assuming a flat prior in $\log
\lambda$,\footnote{We use a flat prior that is uniform in $\log \lambda$ 
instead of $\lambda$ because
we do not know the order of magnitude of $\lambda$ a priori.} the evidence
$P(\dataVec|\lambda, \responseSet, \regSet)$ which appeared in equation
(\ref{eq:posterior}) is the quantity to consider for optimising $\lambda$.

Combining and rearranging equations (\ref{eq:likelihood}), (\ref{eq:prior}),
(\ref{eq:posterior}), (\ref{eq:M}), and (\ref{eq:posterior2}), we get
\be
\label{eq:Evid}
P(\dataVec|\lambda, \responseSet, \regSet) = \frac{Z_{\mathrm{M}}(\lambda)}{Z_{\mathrm{D}} Z_{\mathrm{S}}(\lambda)}.
\ee
For quadratic functional forms of $E_{\mathrm{S}}(\mathbfit{s})$ with
$\mathbfit{s}_{\mathrm{reg}}=\mathbf{0}$, we have 
\be
\label{eq:ZS}
Z_{\mathrm{S}}(\lambda)=e^{-\lambda E_{\mathrm{S}}(\mathbf{0})} \left( \frac{2 \pi}{\lambda} \right)^{N_{\rm s}/2} (\det \mathbfss{C})^{-1/2}, 
\ee
\be
\label{eq:ZM}
Z_{\mathrm{M}}(\lambda)=e^{-M(\mathbfit{s}_{\mathrm{MP}})} (2\pi)^{N_{\rm s}/2} (\det \mathbfss{A})^{-1/2},
\ee
and recall
\be
\label{eq:ZD}
Z_{\mathrm{D}} = (2\pi)^{N_{\rm d}/2} (\det \mathbfss{C}_{\mathrm{D}})^{1/2}.
\ee
Remembering that optimising a function is equivalent to optimising the
logarithm of that function, we will work with $\log P(\dataVec|\lambda,
\responseSet, \regSet)$ to simplify some of the terms.  Recalling that
$\mathbfit{s}_{\mathrm{reg}}=\mathbf{0}$, by combining and
simplifying equations (\ref{eq:Evid}) to (\ref{eq:ZD}), we have 
\bea
\label{eq:EvidFull}
\log P(\dataVec|\lambda, \responseSet, \regSet) &=& -\lambda E_{\mathrm{S}}(\mathbfit{s}_{\mathrm{MP}}) - E_{\mathrm{D}}(\mathbfit{s}_{\mathrm{MP}})  
\nonumber \\ & & - \frac{1}{2}\log(\det\mathbfss{A}) +
\frac{N_{\rm s}}{2}\log\lambda + \lambda E_{\mathrm{S}}(\mathbf{0})
\nonumber \\ & &  + \frac{1}{2}\log(\det\mathbfss{C}) -\frac{N_{\rm d}}{2}\log(2\pi)
\nonumber \\ & & + \frac{1}{2}\log(\det \mathbfss{C}_{\mathrm{D}}^{-1}).
\eea
In deriving equation (\ref{eq:EvidFull}) using equation (\ref{eq:ZS}), we implicitly assumed that $\mathbfss{C}$, the Hessian of $E_{\mathrm{S}}$, is non-singular.  The forms of regularisation we will use for gravitational lensing inversion in Section \ref{sect:appgl} have non-singular Hessians so that equation (\ref{eq:EvidFull}) is applicable.  For the cases in which the Hessian is singular (i.e., at least one of the eigenvalues of the Hessian is zero), the prior probability distribution is uniform along the eigen-directions of the Hessian with zero eigenvalues.  The prior probability distribution will need to be renormalised in the construction of the log evidence expression.  The resulting log evidence expression can still be used to determine the optimal $\lambda$ in these cases because only the relative probability is important and this normalising factor of the uniform prior, though infinite, will cancel in the ratios of probabilities.

Solving  $\frac{\mathrm{d}}{\mathrm{d}\log\lambda} \log P(\dataVec|\lambda, \responseSet,
\regSet)=0$,  we get the following equation for the optimal regularisation
constant $\hat{\lambda}$:
\be
\label{eq:OptLam}
2 \hat{\lambda} E_{\mathrm{S}}(\mathbfit{s}_{\mathrm{MP}}) = N_{\rm s} -
\hat{\lambda} \mathrm{Tr} (\mathbfss{A}^{-1} \mathbfss{C}), 
\ee 
where Tr denotes the trace.  Since $\mathbfit{s}_{\mathrm{MP}}$ and
$\mathbfss{A}$ depend on $\lambda$, the above equation (\ref{eq:OptLam}) is
often nonlinear and needs to be solved numerically for $\hat{\lambda}$.

For the reader's convenience, we reproduce the explanation in \citet{M92} of equation (\ref{eq:OptLam}).  The equation is analogous to the (perhaps) 
familiar statement that $\chi^2$ roughly equals the number of degrees
of freedom.  
Focusing on the usual case where $E_{\mathrm{S}}
(\mathbfit{s}_{\mathrm{reg}}=\mathbf{0})=\mathbf{0}$ and transforming
to the basis in which the Hessian of $E_{\mathrm{S}}$ is the identity
(i.e., $\mathbfss{C}=\mathbfss{I}$), the left-hand side of
equation (\ref{eq:OptLam}) becomes $2\lambda E_{\mathrm{S}}(\mathbfit{s}_{\mathrm{MP}})
= \lambda \mathbfit{s}_{\mathrm{MP}}^{T} 
\mathbfit{s}_{\mathrm{MP}}$.
This quantity can be thought of as the ``$\chi^2_{\mathrm{S}}$
of the parameters'' if we associate $\lambda$ with the width ($\sigma_{\mathrm{S}}$) of the
Gaussian prior: $\lambda = 1/\sigma_{\mathrm{S}}^2$. The
left-hand side of equation (\ref{eq:OptLam}) can be viewed as a
measure of the amount of structure introduced by the data in the
parameter distribution (relative to the null distribution of
$\mathbfit{s}_{\mathrm{reg}}=\mathbf{0}$).  Continuing the analogy, 
the right-hand side of
equation (\ref{eq:OptLam}) is a measure of the number of ``good''
parameters (where ``good'' here means well-determined by the data, 
as we explain below).  In the same basis where $\mathbfss{C}=\mathbfss{I}$,
 we can
write the eigenvalues of $\mathbfss{A}
(=\mathbfss{B}+\lambda\mathbfss{C})$ as $\mu_a + \lambda$, where
$\mu_a$ are the eigenvalues of $\mathbfss{B}$ and index $a=1,\ldots,N_{\rm s}$.  In
this basis, the right-hand side, which we denote by $\gamma$, becomes

\be
\label{eq:gamma}
\gamma = N_{\rm s} - \sum_{a=1}^{N_{\rm s}} \frac{\lambda} {\mu_a+\lambda}
    = \sum_{a=1}^{N_{\rm s}} \frac {\mu_a} {\mu_a + \lambda}.
\ee
For each eigenvalue of $\mathbfss{B}$, the fraction
$\frac{\mu_a}{\mu_a + \lambda}$ is a value between 0 and 1, so
$\gamma$ is a value between 0 and $N_{\rm s}$.  If $\mu_a$ is much smaller
than $\lambda$, then the data are not sensitive to changes in the
parameters along the direction of the eigenvector of $\mu_a$.  This
direction contributes little to the value of $\gamma$ with
$\frac{\mu_a}{\mu_a + \lambda} \ll 1$, and thus it does not constitute
as a good parameter.  Similar arguments show that eigendirections with
eigenvalues much greater than $\lambda$ form good parameters.  Therefore
$\gamma$, which is a sum of all the factors $\frac{\mu_a}{\mu_a +
  \lambda}$, is a measure of the effective number of parameters
determined by the data.  Thus, the solution to equation (\ref{eq:OptLam}) is
the optimal $\lambda$ that matches the $\chi^2_{\mathrm{S}}$ of the
parameters to the number of effective parameters.

For a given form of
regularisation $E_{\mathrm{S}}(\mathbfit{s})$, we are letting the 
data decide on the
optimal $\lambda$ by solving equation (\ref{eq:OptLam}).  
Occam's razor is implicit in this evidence optimisation.  For overly-small
values of $\lambda$, the model parameter space is overly-large and Occam's
razor penalises such an overly-powerful model; for overly-large values of
$\lambda$, the model parameter space is restricted to a limited region that
the model can no longer fit to the data.  Somewhere in between the two
extremes is the optimal $\lambda$ that gives a model which fits to the data
without being overly-complex.

There is a shortcut to obtaining an approximate value of the optimal $\lambda$ instead of solving equation (\ref{eq:OptLam}) \citep{Betal98}.  Given that $\gamma$ is a measure of the effective number of parameters, the classical number of degrees of freedom (NDF) should be $N_{\rm d}-\gamma$.  At the optimal $\lambda$, we thus expect $E_{\mathrm{D}}(\mathbfit{s}_{\mathrm{MP}})=\frac{1}{2}\chi^2\sim \frac{1}{2}\left(N_{\rm d}-\gamma\right)$.  Inserting this and the expression of $\lambda E_{\mathrm{S}}(\mathbfit{s}_{\mathrm{MP}})$ from equation (\ref{eq:OptLam}) into equation (\ref{eq:M}), we find that $M(\mathbfit{s}_{\mathrm{MP}}) \sim \frac{1}{2}N_{\rm d}$.  In other words, one can choose the value of $\lambda$ such that $M$ evaluated at the resulting most probable parameters ($\mathbfit{s}_{\mathrm{MP}}$) is equal to half the number of data points.  We emphasise that this will give only an approximate result for the optimal $\lambda$ due to the fuzzy association of NDF with $N_{\rm d}-\gamma$, but it may serve as a useful hack.


\subsubsection{Ranking models}
\label{sect:BayesInf:ModelComp:rankModel}

We can compare the different regularisations $\regSet$ and responses
$\responseSet$ by examining the posterior probability of $\regSet$ and
$\responseSet$:
\be
\label{eq:postAR}
P(\responseSet,\regSet|\dataVec) \propto P(\dataVec|\responseSet,\regSet) P(\responseSet,\regSet).
\ee

If the prior $P(\responseSet,\regSet)$ is flat, then
$P(\dataVec|\responseSet,\regSet)$ can be used to rank the different models
and regularisations.  We can write $P(\dataVec|\responseSet,\regSet)$ as
\be
\label{eq:PDAR}
P(\dataVec|\responseSet,\regSet) = \int P(\dataVec|\responseSet,\regSet,\lambda) P(\lambda) \mathrm{d}\lambda,
\ee
where $P(\dataVec|\responseSet,\regSet,\lambda)$ is precisely the
evidence in equation~(\ref{eq:EvidFull}).
 
As seen in equation~(\ref{eq:PDAR}) above, 
the regularisation constant  $\lambda$ is a nuisance parameter 
which invariably ends up being marginalised over.
We might well expect the corresponding distribution for $\lambda$ to be
sharply peaked, since we expect the value of $\lambda$ to be 
estimable from the data (as
shown in Section \ref{sect:BayesInf:ModelComp:findLambda});
a particular value of $\lambda$ is preferred as a consequence of the balance
between goodness of fit and Occam's razor.  
Consequently, we can approximate $P(\lambda | \dataVec,
\responseSet, \regSet)$ by a delta function centred on the most probable
constant,~$\hat{\lambda}$.  The model-ranking evidence 
$P(\dataVec|\responseSet,\regSet)$ in equation (\ref{eq:PDAR}) can then be
approximated by $P(\dataVec|\responseSet,\regSet,\hat{\lambda})$ in equation
(\ref{eq:EvidFull}). 

The approximation of using equation (\ref{eq:EvidFull}) to rank regularsations is only valid if the Hessians of the different regularising functions are non-singular.  When the Hessian is singular, equation (\ref{eq:EvidFull}) will need to be modified to include a (infinite) normalisation constant that is regularisation dependent.  The constants for different regularisation schemes generally will not cancel when one considers evidence ratios, thus prohibiting one from comparing different regularisation schemes.

One can imagine there being much debate on the form of the prior  
$P(\responseSet,\regSet)$ that should be used.  For example, some success has
been achieved using  maximum entropy methods \citep[e.g.\ ][]{GD78,S89}, whose
prior form enforces positivity in the image and is maximally non-committal
with regard to missing data. One practical problem with using the entropic
prior is its non-linearity. In this work we take a modern Bayesian view and
argue that while we will always have some a priori prejudice about the
reconstructed image (for example, favouring zero flux, or insisting on positive
images), we would do well to try and learn from the data itself, assigning
series of sensible priors and using the evidence to compare
them quantitatively. In this
context, we examine a small number of sensibly chosen priors (regularisation
schemes), and compute the 
evidence for each.  We do not exhaustively seek the prior that maximizes the
evidence, noting that this will change from object to object, and observation to 
observation. What we do provide is the mechanism by which prior forms can be
compared, and demonstrate that good quality reconstructions can be obtained by
optimising over our set of candidate priors.  In Section \ref{sect:appgl:reg}, we discuss the various forms of prior that have been used in strong gravitational lensing.


\section{Application to gravitational lensing}
\label{sect:appgl}

We apply the Bayesian formalism developed in the previous section to source
inversions in strong gravitational lensing.  The process of finding the
best-fitting pixellated source brightness distribution given a lens potential model and
an observed image has been studied by, for examples, \citet{W96}, \citet{WD03}, \citet{TK04}, \citet{K05},
\citet{DW05} and \citet{BL05}.  The authors regularised the source inversion
in order to obtain a smooth (physical) source intensity distribution.  The forms of regularisation used in this paper are addressed in detail in Appendix \ref{app:regForms}.  In Section \ref{sect:appgl:reg}, we describe the Bayesian analysis of source inversions in gravitational lensing. Sections \ref{sect:appgl:demo1} and \ref{sect:appgl:demo2} are two examples illustrating regularised source inversions.  In both examples, we use simulated data to demonstrate for the first time the Bayesian technique of quantitatively comparing the different types of regularisation.  Finally, Section \ref{sect:appgl:discussion} contains additional discussions based on the two examples.


\subsection{Regularised source inversion}
\label{sect:appgl:reg}

To describe the regularised source inversion problem, we follow \citet{WD03}
but in the Bayesian language.  Let $\data_j$, where $j=1,\ldots,N_{\rm d}$, be the observed
image intensity value at each pixel $j$ and let $\mathbfss{C}_{\mathrm{D}}$ be
the covariance matrix associated with the image data.  Let $s_i$, where
$i=1,\ldots,N_{\rm s}$, be the source intensity value at each pixel $i$ that we would like
to reconstruct.  For a given lens potential and point spread function (PSF)
model, we can construct the $N_{\rm d}$-by-$N_{\rm s}$ matrix $\responseSet$ that maps a source plane
of unit intensity pixels to the image plane by using the lens equation (a
practical and fast method to compute $\responseSet$ is described in the
appendices of \citet{TK04}, and an alternative method is discussed in \citet{W96}).  We identify $E_{\mathrm{D}}$ with
$\frac{1}{2}\chi^2$ (equation (\ref{eq:ED})) and $E_{\mathrm{S}}$ with the quadratic regularising function
whose form is discussed in detail in Appendix \ref{app:regForms}.  The definitions and
notations in our regularised source inversion problem are thus identical to
the Bayesian analysis in Section \ref{sect:BayesInf} with data $\dataVec$ and mapping matrix (response function) 
$\responseSet$.  Therefore, all equations in Section \ref{sect:BayesInf} are
immediately applicable to this source inversion problem, for example the
 most probable
(regularised) source intensity is given by equation (\ref{eq:sMP}).
We take as estimates of the 1-$\sigma$ uncertainty on 
each pixel
value the square root of the corresponding diagonal element of the source covariance 
matrix given by
\be
\label{eq:sCov}
\mathbfss{C}_{\mathrm{S}} = \mathbfss{A}^{-1}
\ee
(here and below, subscript S indicates ``source''), where $\mathbfss{A}$ is the Hessian defined in
Section \ref{sect:BayesInf:ModelFit}.  Equation (\ref{eq:sCov}) differs from the source covariance
matrix used by \citet{WD03}.  We refer the reader to Appendix \ref{app:SrCovM} for details on
the difference.

In summary, to find the most probable source given an image (data) $\dataVec$,
a lens and PSF model $\responseSet$ and a form of regularisation $\regSet$,
the three steps are: (i) find the most likely source intensity,
$\mathbfit{s}_{\mathrm{ML}}$ (the unregularised source inversion with
$\lambda=0$); (ii) solve equation (\ref{eq:OptLam}) for the optimal $\lambda$
of the particular form of regularisation, where $\mathbfit{s}_{\mathrm{MP}}$
is given by equation (\ref{eq:sMP}); (iii) use equations (\ref{eq:sMP}) and
(\ref{eq:sCov}) to compute the most probable source intensity and its
1-$\sigma$ error with the optimal $\lambda$ from step (ii).

Having found a recipe to compute the optimal $\lambda$ and the most probable
inverted source intensity $\mathbfit{s}_{\mathrm{MP}}$ for a given form of
regularisation $\regSet$ and a lens and PSF model $\responseSet$, we can rank the
different forms of regularisation.  For a given potential and PSF model
$\responseSet$, we can compare the different forms of regularisation by
assuming the prior on regularisation $\regSet$ to be flat and using
equations (\ref{eq:postAR}), (\ref{eq:PDAR}), and (\ref{eq:EvidFull}) to
evaluate $P(\responseSet,\regSet|\dataVec)$.  

In this paper, we consider three quadratic functional forms of
regularisation: zeroth order, gradient, and curvature (see Appendix
\ref{app:regForms} for details).  These were used in \citet{WD03} and \citet{K05}.  The zeroth order regularisation
tries to suppress the noise in the reconstructed source brightness
distribution as a way to impose smoothness by minimizing the source
intensity at each pixel.  The gradient regularisation tries to
minimize the gradient of the source distribution, which is equivalent
to minimizing the difference in the source intensities between
adjacent pixels.  Finally, the curvature regularisation minimizes the
curvature in the source brightness distribution.  The two examples in
the following subsections apply the three forms of regularisation to the inversion of
simulated data to demonstrate the Bayesian regularised source
inversion technique.

Our choice of using quadratic functional forms of the prior is
encouraged by the resulting linearity in the inversion.  The linearity
permits fast computation of the maximisation of the posterior without
the risk of being trapped in a local maximum during the optimisation
process.  However, the quadratic functional forms may not be the most
physically motivated.  For example, positive and negative values of the source
intensity pixels are equally preferred, even though we know that
intensities must be positive.  \citet{W96} and \citet{WWLH05} used
maximum entropy methods that enforced positivity on the source
brightness distribution.  Such forms of the prior would help confine
the parameter space of the source distribution and result in a perhaps
more acceptable reconstruction.  The disadvantage of using the entropic
prior is its resulting non-linear inversion, though we emphasise that
Bayesian analysis can still be applied to these situations to rank
models.  Another example is \citet{BL05} who used priors suited for
astronomical images that are mostly blank.  This form of prior also
led to a non-linear system.  In the following sections, we merely
focus on quadratic forms of the prior because (i) it is computational
efficiency, and (ii) we could obtain good quality reconstruction
without considering more complex regularisation schemes.


\subsection{Demonstration 1: Gaussian Sources}
\label{sect:appgl:demo1}


\subsubsection{Simulated data}
\label{sect:appgl:demo1:simData}

\begin{figure*}
\vspace{0.1in}
\includegraphics[width=170mm]{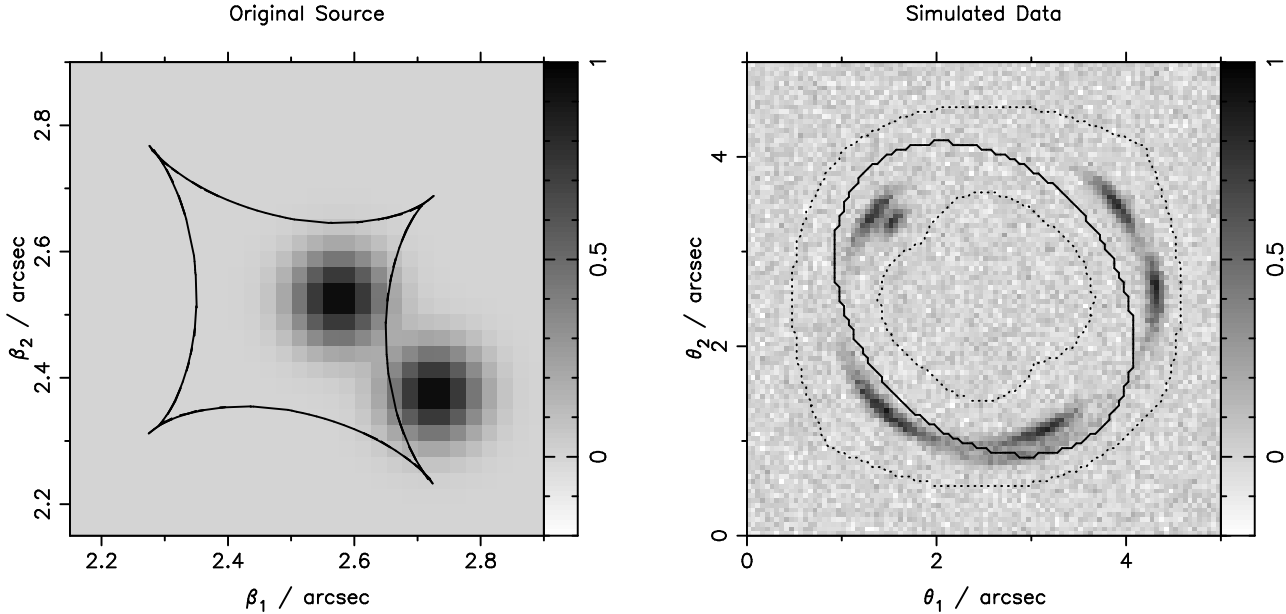}

\caption{\label{fig:simData14} Left-hand panel: The simulated Gaussian sources with peak
intensities of 1.0 and FWHM of $0.05''$, shown with the astroid caustic curve
of the SIE potential.  Right-hand panel: The simulated image of the Gaussian 
sources (after
convolution with Gaussian PSF and addition of noise, as described in the text).
The solid line is the
critical curve of the SIE potential, and the dotted lines mark the annular
region where the source grid maps using the mapping matrix $\responseSet$. }

\end{figure*}
 
As the first example to demonstrate the Bayesian approach to source inversion, we use the same lens potential and source brightness distribution as that in \citet{WD03}.  The lens is a
singular isothermal ellipsoid (SIE) at a redshift of $z_{\rm d}=0.3$ with
one-dimensional velocity dispersion of $260 \hspace{0.05cm} \mathrm{km \hspace{0.05cm} s^{-1}}$, axis ratio of
$0.75$, and semi-major axis position angle of $40$ degrees (from vertical in
counterclockwise direction).  We use \citet*{KSB94} for the SIE model.  We
assume a flat $\Lambda$-CDM universe with cosmological parameters of $\Omega_m
= 0.3$ and $\Omega_{\Lambda}=0.7$.  The image pixels are square and have sizes
$0.05''$ in each direction.  We use $100\times100$ image pixels ($N_{\rm d}=10000$)
in the simulated data. 

We model the source as having two identical Gaussians with variance $0.05''$
and peak intensity of 1.0 in arbitrary units.  The source redshift is
$z_{\rm s}=3.0$.  We set the source pixels to be half the size of the image pixels
($0.025''$) and have $30\times30$ source pixels ($N_{\rm s}=900$). 
Fig.\ref{fig:simData14} shows the source in the left-hand panel with the caustic
curve of the SIE potential.  One of the Gaussians is located within the astroid
caustic and the other is centred outside the caustic.  

To obtain the simulated data, we use the SIE lens model and the lens equation
to map the source intensity to the image plane.  We then convolve the
resulting image with a Gaussian PSF whose FWHM is $0.08''$ and add  
Gaussian noise of variance $0.067$ to the convolved image.  For
simplicity, the noise is uncorrelated, which is a good approximation 
to realistic noise with minimal charge transfer and drizzling.  
The right-hand panel of
Fig.~\ref{fig:simData14} shows the simulated data with the critical curve of
the SIE model.


\subsubsection{Most likely inverted source}
\label{sect:appgl:demo1:mostLikely}

We use the original SIE potential, PSF and Gaussian noise models of the simulated data for the
source inversion to demonstrate the technique.  

The appendices of \citet{TK04} describe a computationally efficient method to
construct the $\responseSet$ matrix.  Following the method, we discretize
the SIE potential to the $100\times100$ grid and model the PSF on a $5\times5$
grid (which is a sufficient size since the $5\times5$ grid centred on the
Gaussian PSF of FWHM 0.08'' contains 99.99 per cent of the total intensity). 
Subsequently, for every image pixel $j$, we use the lens equation to trace to
the source plane labelled by pixels $i$ and interpolate to get the elements of unblurred
$\responseSet$.  Lastly, we multiply the unblurred $\responseSet$ by the
blurring (convolution) operator constructed from the $5\times5$ PSF model to
get the full $\responseSet$ matrix.  With $j=1,\ldots,N_{\rm d}$ and $i=1,\ldots,N_{\rm s}$, the
matrix $\responseSet$ is large ($10000\times900$) but fortunately sparse.

In the right-hand panel of Fig.~\ref{fig:simData14}, the dotted lines on the
simulated data mark an annular region where the image pixels map to the finite
source plane.  In other words, the image pixels within the dotted annulus
correspond to the non-empty rows of the $\responseSet$ matrix.  The annular
region thus marks the set of data that will be used for the source inversion
process.

With the $\responseSet$ matrix and the data of simulated image intensities
in the annulus, we can construct matrix $\mathbfss{F}$ and vector
$\mathbfit{D}$ using equations (\ref{eq:F}) and (\ref{eq:D})\footnote{The
summations associated with the matrix multiplications 
in equations (\ref{eq:F}) and (\ref{eq:D}) are now summed over the pixels in
the annulus instead of all the pixels on the image plane.} for the
unregularised inversion (the most likely source intensity, in Bayesian
language).  We use UMFPACK\footnote{a sparse matrix package developed by
Timothy A. Davis, University of Florida} for sparse matrix inversions and
determinant calculations.  We compute the inverse of the matrix $\mathbfss{F}$
and apply equation (\ref{eq:sML}) to get the most likely source intensity. 
Using UMFPACK, the computation time for the inversion of $\mathbfss{F}$, a
$900\times900$ matrix in this example, is only $\sim 20$ seconds on a 3.6 GHz
CPU.  Setting $\lambda=0$ (implicit in $\mathbfss{A}$) in equation
(\ref{eq:sCov}), we obtain the covariance matrix of the inverted source
intensity and hence the 1-$\sigma$ error and the signal-to-noise ratio.  

\begin{figure*}
\includegraphics[width=170mm]{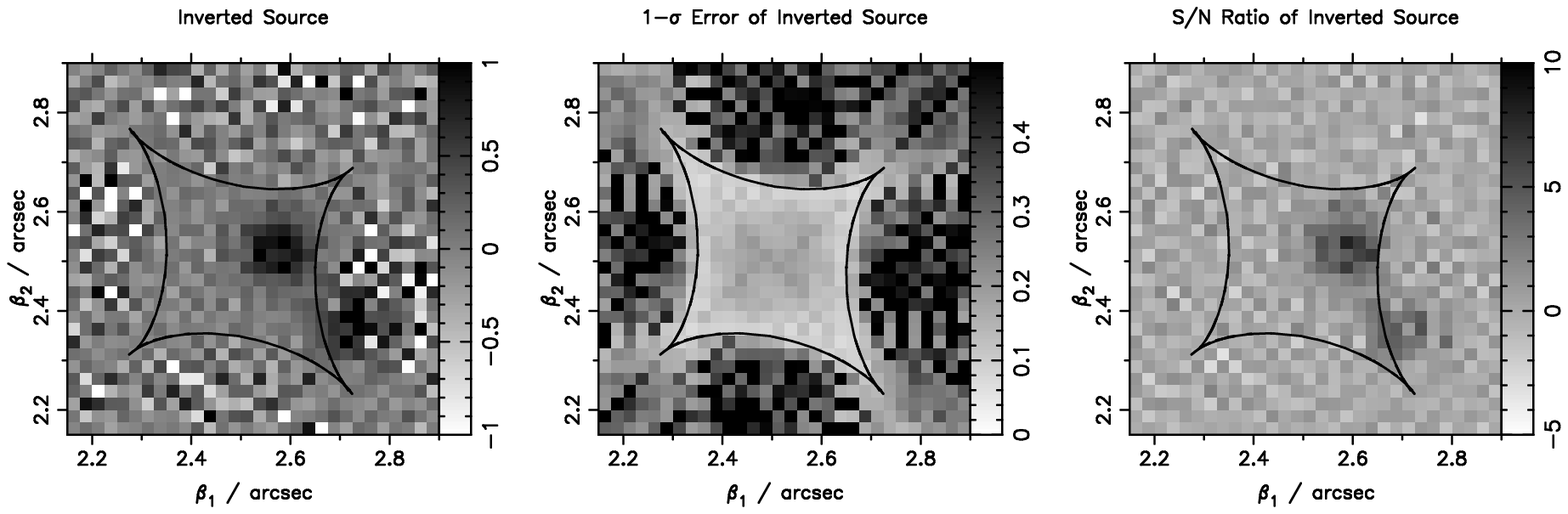}
\vspace{0.in}
\includegraphics[width=170mm]{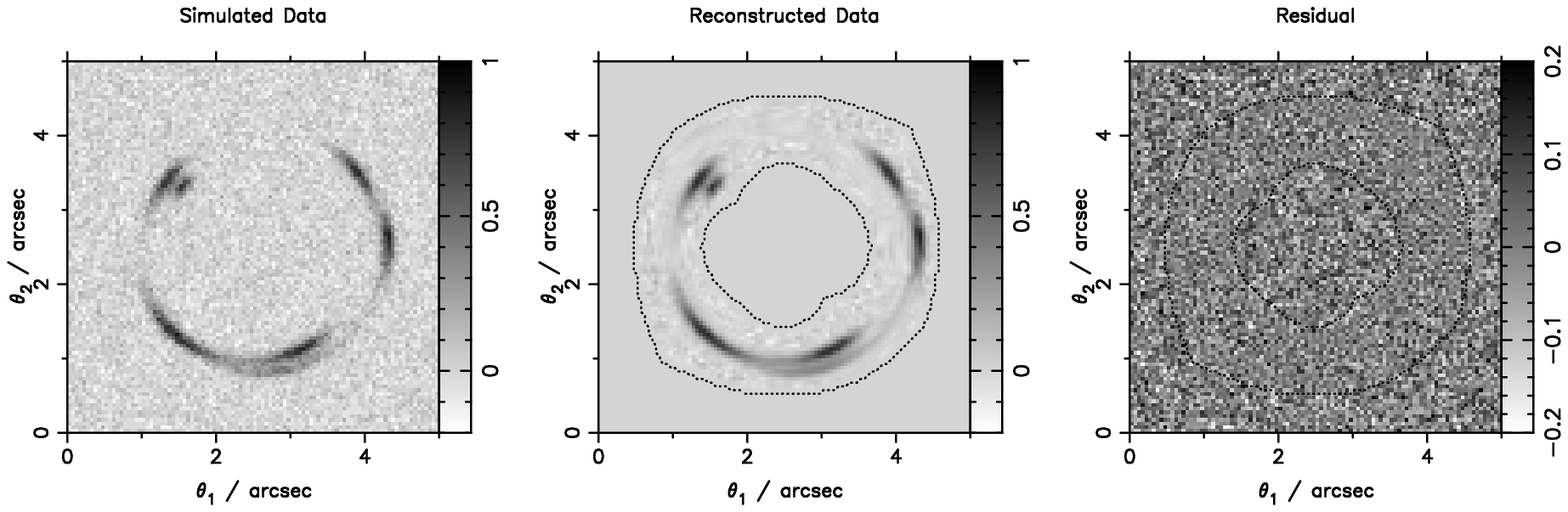}

\caption{\label{fig:NRSr14} Unregularised inversion of Gaussian sources.  Top left-hand panel:
the most likely reconstructed source intensity distribution.  The intensities
outside the caustic curve of the potential model are not well-reconstructed
due to fewer constraints (lower image multiplicities) outside the caustic
curve.  Top middle panel: the 1-$\sigma$ error of the inverted source intensity.  The
error is smaller inside the caustics due to additional multiple image
constraints.  Top right-hand panel: the signal-to-noise ratio of the inverted source
intensity.  The presence of the Gaussian sources is clear in this panel even
though the reconstruction in the top left-hand panel is noisy. 
Bottom left-hand panel: the simulated data.  Bottom middle panel: the reconstructed image using the
most likely reconstructed source (top left-hand panel) and the
$\responseSet$ matrix from the potential and PSF models.  Reconstructed data
is confined to an annular region that maps on to the source plane.  Bottom right-hand panel: the
residual image obtained by subtracting the bottom middle panel from the bottom left-hand panel. 
The interior of the annular region is less noisy than the exterior, indicating
that the unregularised reconstructed source is fitting to the noise in the
simulated data.}

\end{figure*}

The top row of Fig.~\ref{fig:NRSr14} shows the unregularised inverted source intensity in the
left-hand panel, the 1-$\sigma$ error of the intensity in the middle panel, and the
signal-to-noise ratio in the right-hand panel.  The unregularised inverted source
intensity is smoother inside than outside the caustic curve because the source
pixels within the caustic have additional constraints due to higher image
multiplicities.  The higher image multiplicities also explain the lower
magnitude of the 1-$\sigma$ error inside the caustic curve.  Despite the noisy
reconstruction especially outside the caustic curve, the two Gaussian sources
have significant signal-to-noise in the right-hand panel.  These results agree with
Fig.2 in \citet{WD03}.  

The bottom row of Fig.~\ref{fig:NRSr14} shows the simulated data in the left-hand panel (from
Fig.~\ref{fig:simData14} for comparison purposes), the reconstructed data
(from the most likely inverted source in the top left-hand panel and the
$\responseSet$ matrix) in the middle panel, and the residual (the difference
between the simulated and reconstructed data) in the right-hand panel.  The annular
region containing the data used for inversion is marked by dotted lines in the
reconstructed and residual images.  Visual inspection of the residual image
shows that pixels inside the annulus are slightly less noisy than those
outside.  This is due to over-fitting with the unregularised inversion.  As we
will see in the next subsection, Occam's razor that is incorporated in the
Bayesian analysis will penalise such overly-powerful models.

\subsubsection{Most probable inverted source}
\label{sect:appgl:demo1:mostProbable}

Having obtained the most likely inverted source, we can calculate the most
probable source of a given form of regularisation with a given value of the
regularisation constant $\lambda$ using equation (\ref{eq:sMP}).  In the
remainder of this section, we focus on the three forms of regularisation
(zeroth-order, gradient, and curvature) discussed in Appendix \ref{app:regForms}.  For each form
of regularisation, we numerically solve equation (\ref{eq:OptLam}) for the
optimal value of regularisation constant $\lambda$ using equation
(\ref{eq:sMP}) for the values of $\mathbfit{s}_{\mathrm{MP}}$.  Table
\ref{tab:OptLambda} shows the optimal regularisation constant, $\hat \lambda$,
for each of the three forms of regularisation.  The table also includes the
value of the evidence in equation (\ref{eq:EvidFull}) evaluated at $\hat
\lambda$, which is needed for ranking the different forms of regularisation in
the next subsection.

\begin{table}
\caption{\label{tab:OptLambda} The optimal regularisation constant for each of
the three forms of regularisation for the inversion of two Gaussian sources. 
The log evidence, $\gamma$ (the right hand side of equation (\ref{eq:OptLam})), and the $\chi^2$   
evaluated at the optimal regularisation constant are also
listed.  The number of data pixels in the annulus for inversion, $N_{\rm annulus}$, and three possible forms of constructing the reduced $\chi^2$ are shown.}  

\begin{tabular}{|c|c|c|c|}
\hline
Regularisation  & zeroth-order & gradient & curvature \\
\hline 
$\hat{\lambda}$ & 17.7 & 34.2 & 68.5 \\
$\log P(\dataVec|\hat{\lambda}, \responseSet, \regSet)$ & 5086 & 5367
& 5410 \\
$\gamma = N_{\rm s} - \hat{\lambda} \mathrm{Tr} (\mathbfss{A}^{-1} \mathbfss{C})$ & 536 & 287 & 177 \\
$\chi^2=2E_{\mathrm{D}}$ & 3583 & 3856 & 4019\\
$N_{\rm annulus}$ & 4325 & 4325 & 4325 \\
$\chi^2/N_{\rm annulus}$ & 0.83 & 0.89 & 0.93\\
$\chi^2/(N_{\rm annulus}-N_{\rm s})$ & 1.05 & 1.12 & 1.17 \\
$\chi^2/(N_{\rm annulus}-\gamma)$ & 0.95 & 0.95 & 0.97 \\
\hline
\end{tabular}
\end{table}

\begin{figure}
\vspace{0.3in}
\includegraphics[width=84mm]{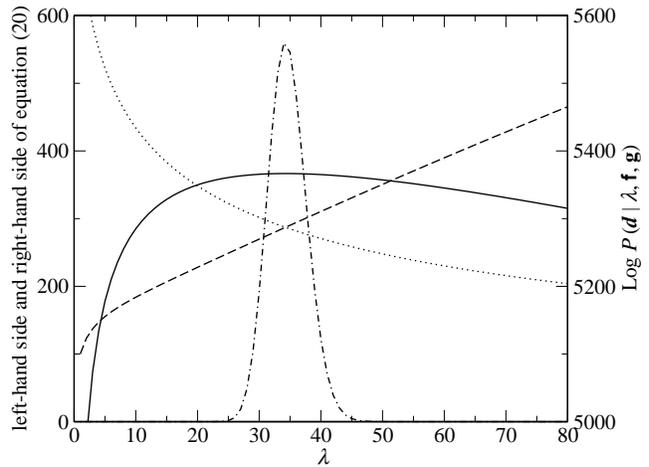}

\caption{\label{fig:BayesDemo} To demonstrate the $\lambda$
optimisation process, equations (\ref{eq:EvidFull}) and
(\ref{eq:OptLam}) are plotted as functions of $\lambda$ for the
gradient regularisation.  The left-hand side and right-hand side of
equation (\ref{eq:OptLam}) are in dashed lines and dotted lines,
respectively.  The log evidence in equation (\ref{eq:EvidFull}) is
shown in solid lines.  The evidence, which as been rescaled to fit on
the graph, is in dot-dashed lines.  The left and right vertical axes
are for equation (\ref{eq:OptLam}) and (\ref{eq:EvidFull}),
respectively.  The crossing point of the left-hand side and right-hand
side of equation (\ref{eq:OptLam}) gives the optimal $\hat \lambda$,
the position where the log evidence (hence evidence) obtains its
maximum.}
\end{figure}

Fig.~\ref{fig:BayesDemo} verifies the optimisation results for the
gradient form of regularisation.  The
evidence in dot-dashed lines (rescaled) is indeed a sharply-peaked function of $\lambda$, justifying the
delta-function approximation; the optimal regularisation constant $\hat
\lambda =34.2$ (listed in Table \ref{tab:OptLambda}) is marked by the crossing
point of the dashed and dotted lines, demonstrating the balance between
goodness of fit and simplicity of model that maximising the evidence
achieves.  The plots of equations (\ref{eq:OptLam}) and (\ref{eq:EvidFull})
for zeroth-order and curvature regularisations look similar to
Fig.~\ref{fig:BayesDemo} and are thus not shown.

In Table \ref{tab:OptLambda}, we constructed three reduced $\chi^2$ using the NDF as $N_{\rm annulus}$, $N_{\rm annulus}-N_{\rm s}$, or $N_{\rm annulus}-\gamma$, where $N_{\rm annulus}$ is the number of data pixels used in the inversion and recall $N_{\rm s}$ is the number of source pixels reconstructed.  In each of the three forms of regularisation, the reduced $\chi^2$ with $\mathrm{NDF}=N_{\rm annulus}-\gamma$ is closest to 1.0, which is the criterion commonly used to determine the goodness of fit.  This supports our interpretation of the $\gamma$, the right-hand side of equation (\ref{eq:OptLam}), as the number of ``good'' parameters determined by the data.  The values of the reduced $\chi^2$ is not strictly 1.0 because Bayesian analysis determines the optimal $\lambda$ by maximizing the evidence instead of setting the reduced $\chi^2$ to 1.0.

\begin{figure*}
\includegraphics[width=170mm]{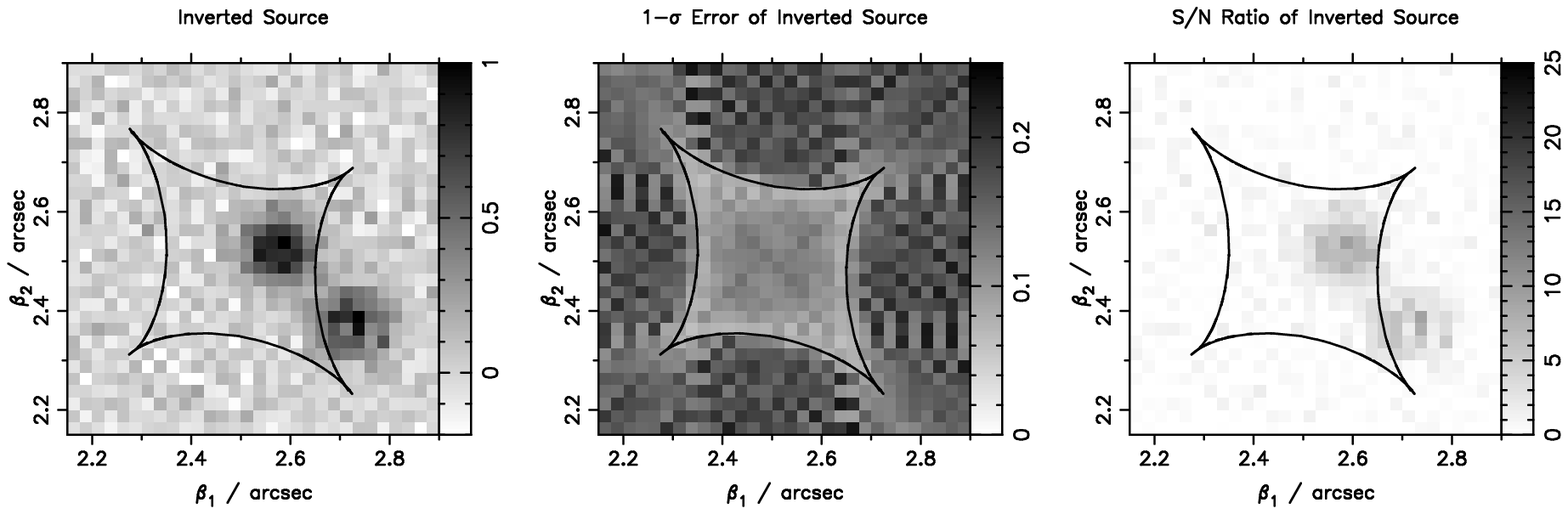}
\vspace{0.1in}
\includegraphics[width=170mm]{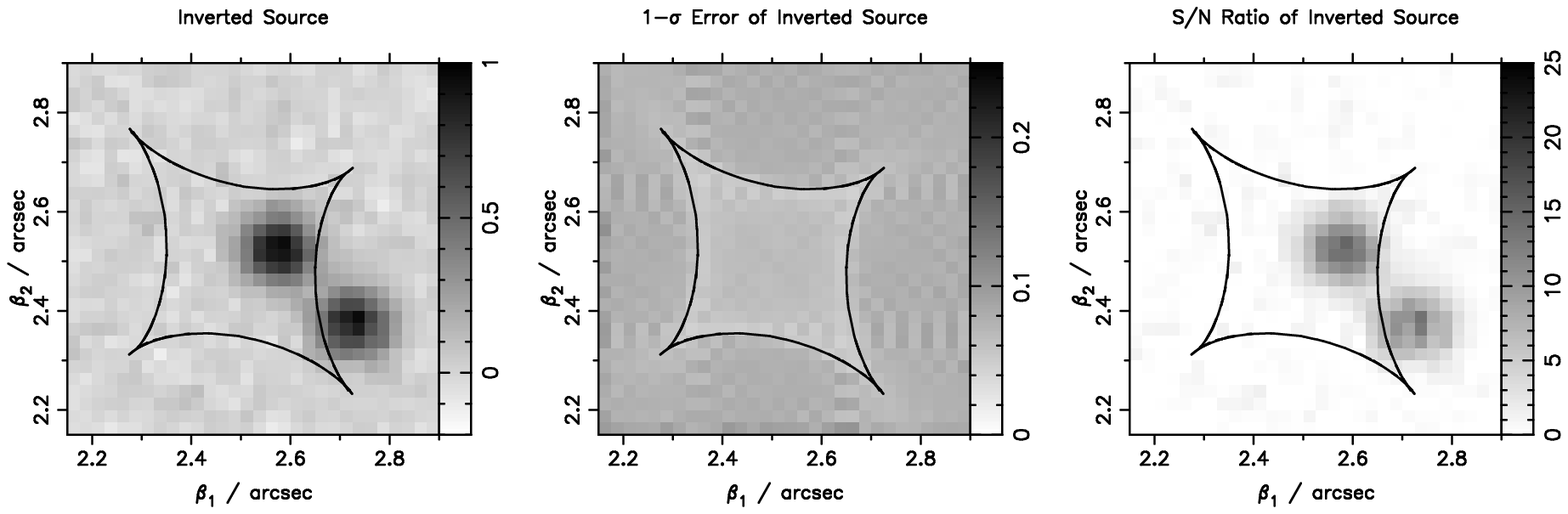}
\vspace{0.1in}
\includegraphics[width=170mm]{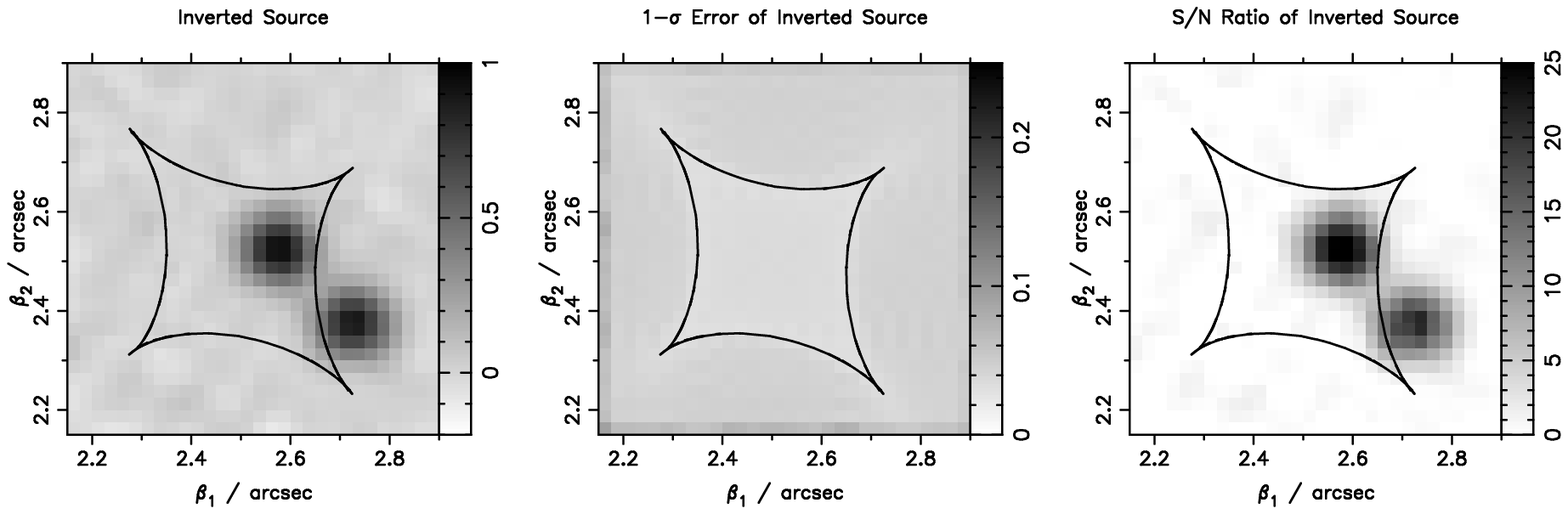}

\caption{\label{fig:RSr14}  The regularised source inversions of Gaussian
sources with zeroth-order, gradient and curvature regularisations.  Top row, from
left to right: most probable inverted source, the 1-$\sigma$ error, and the
signal-to-noise ratio with zeroth-order regularisation.  Middle row, from left to
right: same as top row but with gradient regularisation.  Bottom row, from
left to right: same as top row but with curvature regularisation.  The panels
in each column are plotted on the same scales for comparison among the
different forms of regularisation.}

\end{figure*}

For each of the three forms of regularisation and its optimal regularisation
constant listed in Table \ref{tab:OptLambda}, we use equations (\ref{eq:sMP})
and (\ref{eq:sCov}) to obtain the most probable source intensity and its
1-$\sigma$ error.  Fig.\ref{fig:RSr14} shows the most probable source
intensity (left-hand panels), the 1-$\sigma$ error (middle panels), and the
signal-to-noise ratio (right-hand panels) for zeroth-order (top row), gradient (middle
row) and curvature (bottom row) regularisations.  The panels in each column
are plotted on the same scales in order to compare the different forms of
regularisation.  The regularised inverted sources in the left-hand panels clearly
show the two Gaussians for all three regularisations.  Curvature
regularisation results in a smoother source reconstruction than gradient
regularisation which in turn gives smoother source intensities than zeroth-order
regularisation.  The 1-$\sigma$ errors in the middle column also indicates the
increase in the smoothness of the source from zeroth-order to gradient to
curvature regularisation due to a decrease in the error.  This smoothness
behaviour agrees with our claim in Appendix \ref{app:regForms} that regularisations
associated with higher derivatives in general result in smoother source
reconstructions.  Since the error in the middle column decreases from the top
to the bottom panel, the signal-to-noise of the source reconstruction
increases in that order.  Looking closely at the 1-$\sigma$ error in the
middle column for gradient and curvature regularisations, the pixels in the
left and bottom borders have larger error values.  This can be explained by
the explicit forms of regularisation in equations (\ref{eq:ESgrad}) and
(\ref{eq:EScurv}).  The pixels at the bottom and left borders are only
constrained by their values relative to their neighbours, whereas the pixels
at the top and right borders have additional constraints on their values
directly (last two terms in the equations).  Visually, we observe that the
source reconstruction with curvature regularisation matches the original
source in Fig.~\ref{fig:simData14} the best.  In the next subsection, we will
quantitatively justify that curvature regularisation is preferred over
gradient and zeroth-order regularisations in this example with two Gaussian
sources.

\begin{figure*}
\vspace{0.0in}
\includegraphics[width=170mm]{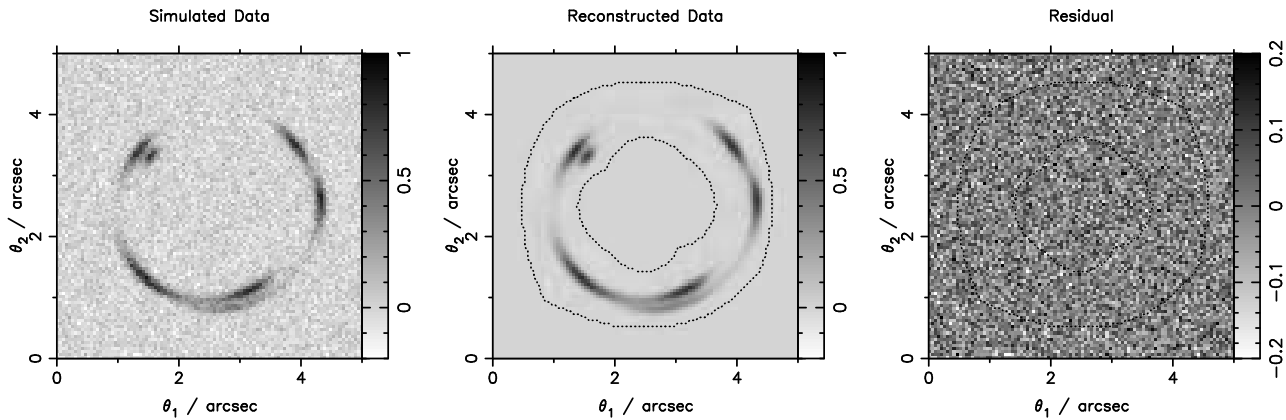}

\caption{\label{fig:RRes14} The image residual for curvature regularised
source inversion with Gaussian sources.  From left to right: simulated data,
reconstructed data using the corresponding most probable inverted source in
Fig.~\ref{fig:RSr14}, and the residual equalling the difference between
simulated and reconstructed data.  The reconstructed data is restricted to the
annulus marked by dotted lines that is mapped from the finite source grid
using $\responseSet$.  The noise in the residual image is more uniform
compared to that of the unregularised inversion in Fig.~\ref{fig:NRSr14}.}

\end{figure*}

In Fig.~\ref{fig:RRes14}, we show the reconstructed image and the image
residual for the most probable inverted source with curvature regularisation. 
We omit the analogous figures for zeroth-order and gradient regularisations
because they look very similar to Fig.~\ref{fig:RRes14}.  The left-hand panel is
the simulated data in Fig.~\ref{fig:simData14} that is shown for convenience
for comparing to the reconstructed data.  The middle panel is the
reconstructed data obtained by multiplying the corresponding regularised
inverted source in Fig.~\ref{fig:RSr14} by the $\responseSet$ mapping matrix
(only the pixels within the annulus [dotted lines] are reconstructed due to
the finite source grid and PSF).  The right-hand panel is the residual image, which
is the difference between the simulated and the reconstructed data.  The
slight difference among the reconstructed data of the three forms of
regularisations is the amount of noise.  Since the most probable inverted
source gets less noisy from zeroth-order to gradient to curvature regularisation,
the reconstructed data also gets less noisy in that order.  The residual
images of all three forms of regularisation look almost identical and
match the input (uniform Gaussian) noise, a sign of proper source reconstruction.  

\begin{figure*}
\vspace{0.0in}
\includegraphics[width=170mm]{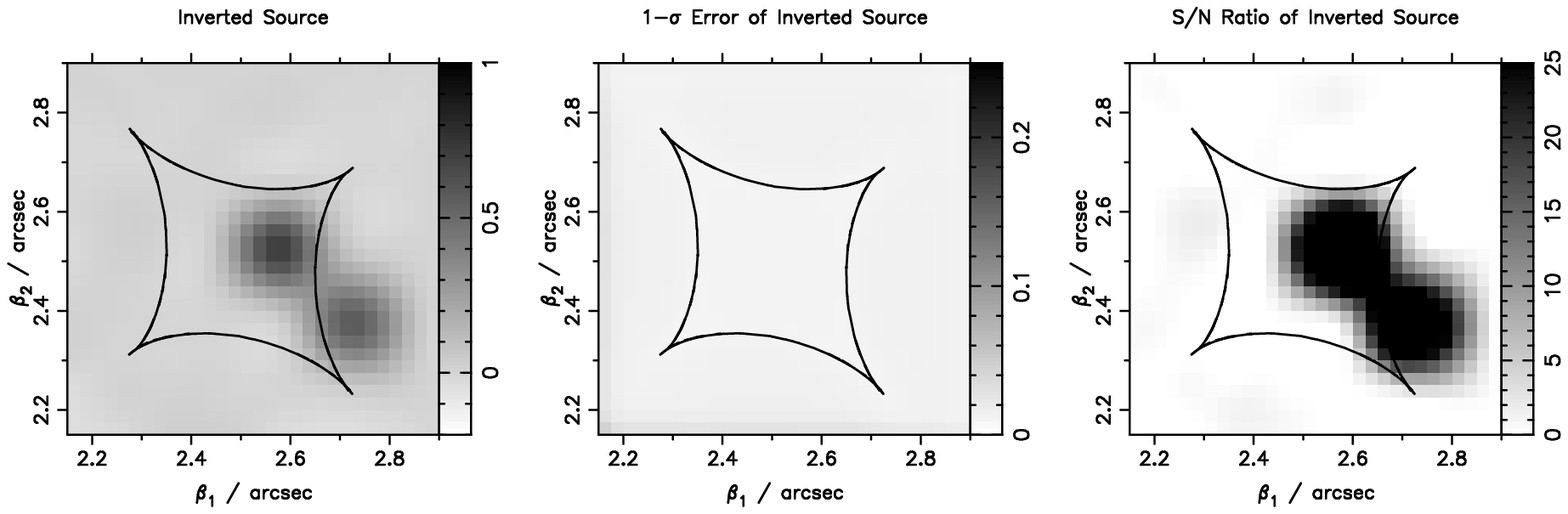}
\vspace{0.0in}
\includegraphics[width=170mm]{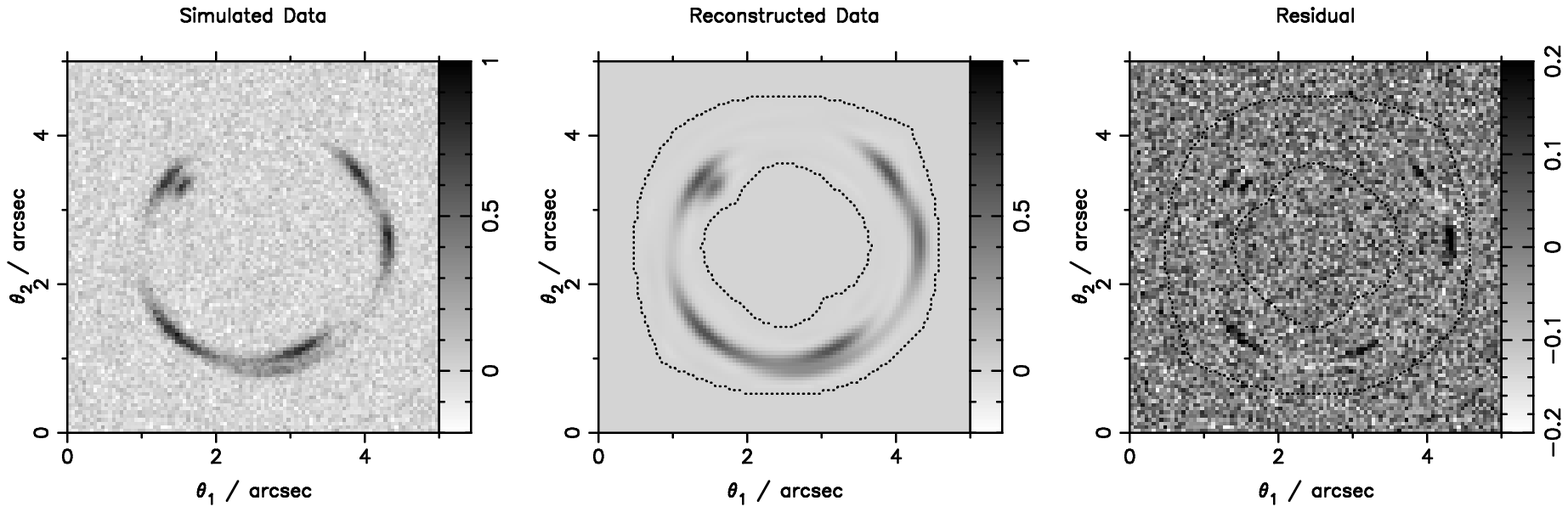}

\caption{\label{fig:CR2000Sr14} Overly-regularised source inversion of
Gaussian sources using curvature regularisation with $\lambda=2000$.  Top row: 
the overly-regularised source shows smearing of the original two Gaussians (left-hand panel),
 the 1-$\sigma$ error of the source intensity (middle panel), and the
signal-to-noise ratio (right-hand panel).  
Bottom row:  simulated data (left-hand panel),  reconstructed data using the
reconstructed source in the top left-hand panel and the $\responseSet$
mapping matrix (middle panel), and the image residual showing arc features due to the
overly-regularised inverted source (right-hand panel).}

\end{figure*}

In contrast to the residual image for the unregularised case in
Fig.~\ref{fig:NRSr14}, the noise in the residual image in
Fig.~\ref{fig:RRes14} is more uniform.  This is Occam's razor in action - the
presence of regularisation prevents the over-fitting to the noise within the
annulus.  For each form of regularisation, the value of $\hat \lambda$ (Table
\ref{tab:OptLambda}) is optimal since it leads to
the residual image in Fig.~\ref{fig:RRes14} having the input noise, which is uniform
Gaussian noise in our example.  If we over-regularise (i.e.,
use overly large $\lambda$), then we expect the model to no longer fit to the
data.  This is shown in Fig.~\ref{fig:CR2000Sr14}
which were obtained using curvature regularisation with $\lambda=2000$.  The 
panels in the figure are displayed in the same way as in
Fig.~\ref{fig:NRSr14}.  The inverted source (top left-hand panel) in
Fig.~\ref{fig:CR2000Sr14} shows the smearing of the two Gaussian sources due
to overly-minimized curvature among adjacent pixels.  The resulting residual
image (bottom right-hand panel) in Fig.~\ref{fig:CR2000Sr14} thus shows arc features that
are not fitted by the model.  
However, note that the inferred signal-to-noise ratio in the source plane 
is very high; models that overly-regularise the source intensities give
precise (with small magnitudes for the error) but inaccurate results.  Such
overly-regularised models lead to low values of the evidence, which is the
quantity to consider for the goodness of reconstruction.  
We seek an accurate reconstruction of the source, and a
signal-to-noise ratio that accurately reflects the noise in the data. 
The comparison among the unregularised,
optimally regularised and overly-regularised inversions shows the power of the
Bayesian approach to objectively determine the optimal $\hat \lambda$ (of a
given form of regularisation) that minimizes the residual without fitting to
the noise.  In the next subsection, we will see how Bayesian analysis can also
be used to determine the preferred form of regularisation given the 
selection of regularisations.

\subsubsection{Optimal form of regularisation}
\label{sect:appgl:demo1:OptReg}

In the previous subsection, we showed how Bayesian analysis allowed us to
determine objectively the optimal regularisation constant for a given form of
regularisation by maximizing the evidence in equation (\ref{eq:EvidFull}).  In
this subsection we look for the optimal form of regularisation given
the selection of regularisations.

Since there is no obvious prior on the regularisation, we assume that the
prior on the regularisation is flat.  In this case, the different forms of
regularisation is ranked by the value of $P(\dataVec|\responseSet,\regSet)$ in
equation (\ref{eq:PDAR}).  Since the evidence
$P(\dataVec|\responseSet,\regSet, \lambda)$ is sharply peaked at $\hat
\lambda$ (as seen in Fig.~\ref{fig:BayesDemo}),
$P(\dataVec|\responseSet,\regSet)$ can be 
approximated by $P(\dataVec|\responseSet,\regSet, \hat \lambda)$.
The values of the evidence
$P(\dataVec|\responseSet,\regSet, \hat \lambda)$ in Table \ref{tab:OptLambda}
indicate that the evidence for curvature regularisation is $\sim
\mathrm{e}^{43}$ and $\sim \mathrm{e}^{324}$ higher than that of gradient and
zeroth-order regularisations, respectively.  
  Therefore, curvature
regularisation with the highest evidence is preferred to zeroth-order and
gradient for the two Gaussian sources.  In quantitative terms, curvature
regularisation is $\sim \mathrm{e}^{43}$ more probable than gradient
regularisation, which is $\sim \mathrm{e}^{281}$ more probable than zeroth-order
regularisation.  This agrees with our comment based on Fig.~\ref{fig:RSr14} in
Section \ref{sect:appgl:demo1:mostProbable} that visually, curvature regularisation leads to an inverted
source that best matches the original source of two Gaussians.

The values of the reduced $\chi^2$ using $\mathrm{NDF}=N_{\rm annulus}-\gamma$ in Table \ref{tab:OptLambda} show that curvature regularisation has the highest reduced $\chi^2$ among the three forms of regularisation.  The higher $\chi^2$ value means a higher misfit due to fewer degrees of freedom (with more correlated adjacent pixels) in curvature regularisation.  Nonetheless, the misfit is noise dominated since Fig.~\ref{fig:RRes14} shows uniform residual and the reduced $\chi^2$ is $\sim 1.0$.  Therefore, the evidence optimisation is selecting the simplest model of the three regularisation schemes that fits to the data, enforcing Occam's razor.

For general source brightness distributions, one may expect that curvature regularisation
with its complex structure will always be preferred to the simplistic gradient
and zeroth-order forms of regularisation.  We show that this is not the case by
considering the source inversion of a box source (region of uniform intensity)
and two point sources as our next example.


\subsection{Demonstration 2: box and point sources}
\label{sect:appgl:demo2}


\subsubsection{Simulated data}
\label{sect:appgl:demo2:simData}
To generate the simulated data of the box and point sources, we keep the following
things the same as those in the example of two Gaussian sources: number of
source pixels, source pixel size, number of image pixels, image pixel size,
SIE potential model, and PSF model.  The variance of the uniform uncorrelated Gaussian
noise for the box and point sources is 0.049, which leads to the same
signal-to-noise ratio within the annular region as that in the two Gaussian
sources.  Fig.~\ref{fig:simData17} shows the box source and two point sources
of unit intensities with the caustic curves of the SIE in the left-hand panel, and
the simulated image in the right-hand panel.  

We follow the same procedure as that
in the previous example of two Gaussian sources to obtain the most likely
inverted source, the most probable inverted source of a given form of
regularisation, and the optimal form of regularisation.  Furthermore, we plot
the results in the same format as that in the example of two Gaussian
sources in Section \ref{sect:appgl:demo1}.  


\subsubsection{Most likely inverted source, most probable inverted source, and optimal form of regularisation}
\label{sect:appgl:demo2:allThree}
Figs.~\ref{fig:NRSr17} shows the most likely
inverted source in the top row and the corresponding image residual in the bottom row.  Similar
to Fig.~\ref{fig:NRSr14}, the most likely inverted source in the top left-hand panel of
Fig.~\ref{fig:NRSr17} has poorly constrained pixels outside the caustic curves
due to lower image multiplicities.  The residual image in the bottom right-hand panel of
Fig.~\ref{fig:NRSr17} shows slight over-fitting to the noise inside the
annulus.  

\begin{figure*}
\vspace{0.1in}
\includegraphics[width=170mm]{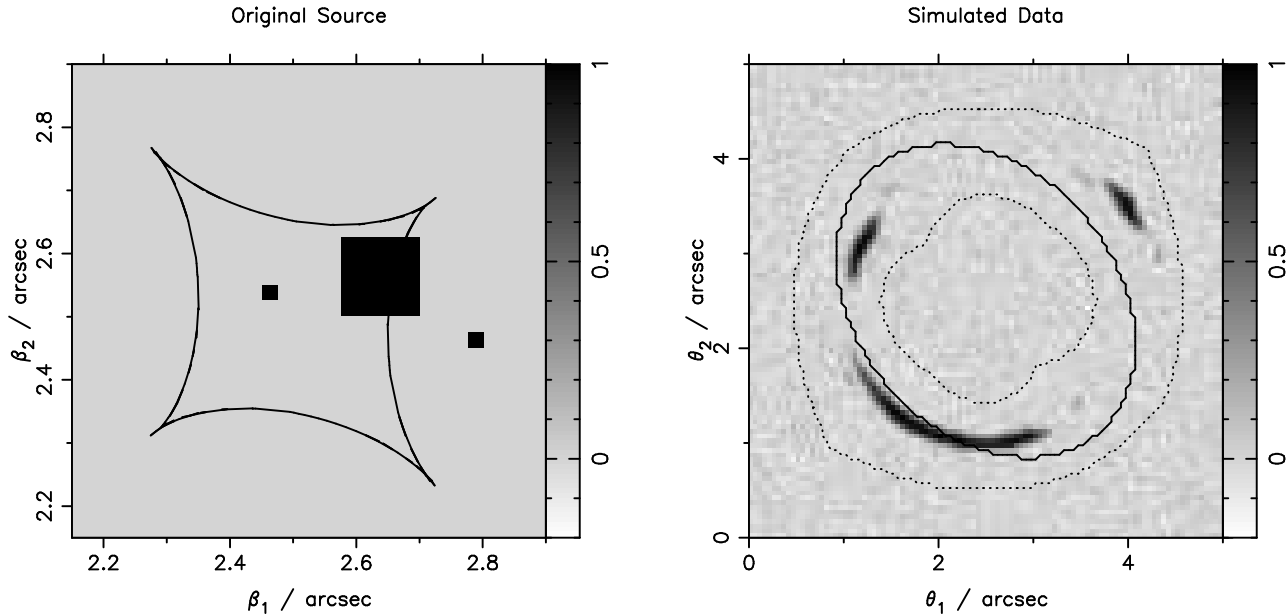}

\caption{\label{fig:simData17} Left-hand panel: The simulated box and point sources with
intensities of 1.0, shown with the astroid caustic curve of the SIE
potential.  Right-hand panel: The simulated image of the box and point sources 
(after convolution with Gaussian PSF and addition of noise as described in the
text).  The solid line is the critical curve
of the SIE potential and the dotted lines mark the annular region where the source
grid maps using the $\responseSet$ mapping matrix. }

\end{figure*}
\begin{figure*}
\vspace{0.0in}
\includegraphics[width=170mm]{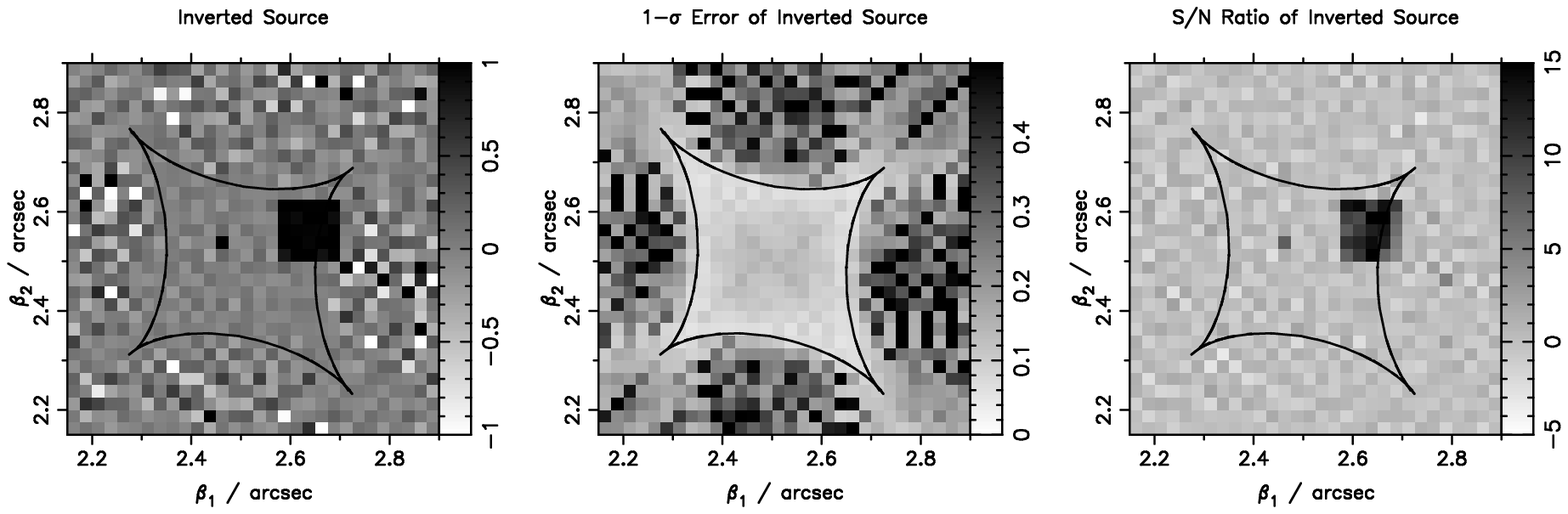}
\vspace{0.0in}
\includegraphics[width=170mm]{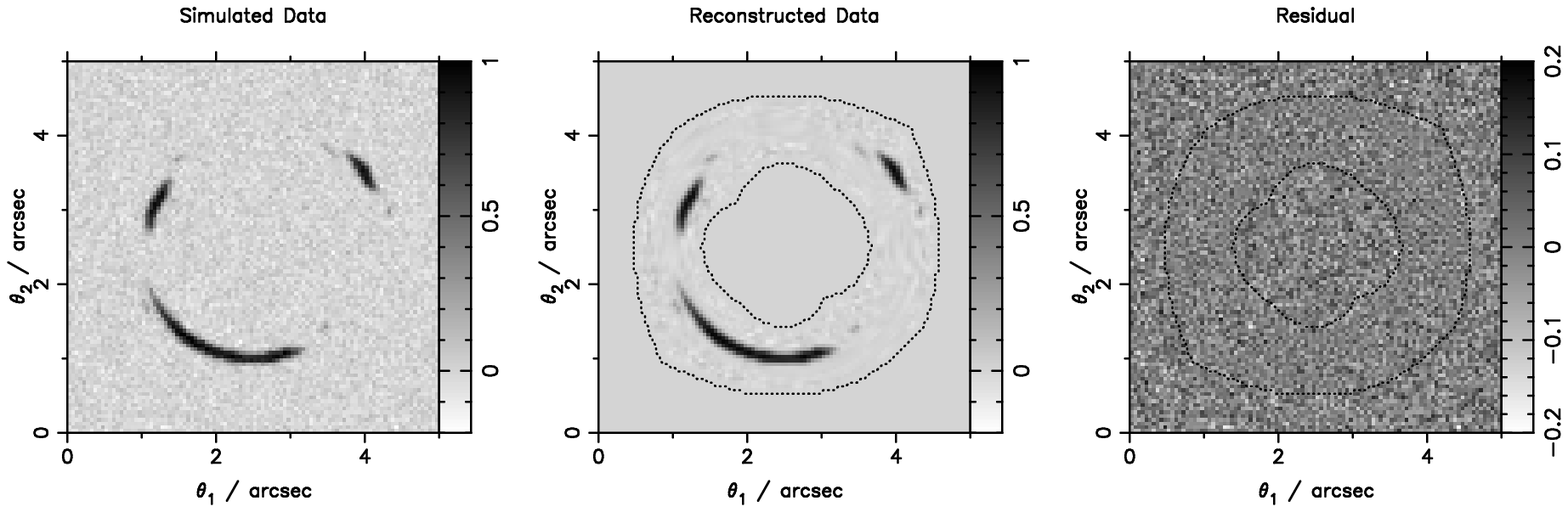}

\caption{\label{fig:NRSr17} Unregularised source inversion of box and point
sources.  Top left-hand panel: the most likely reconstructed source intensity distribution. 
The intensities outside the caustic curve of the potential model are not
well-reconstructed due to fewer constraints (lower image multiplicities)
outside the caustic curve.  Top middle panel: the 1-$\sigma$ error of the inverted
source intensity.  The error is smaller inside the caustics due additional
multiple image constraints.  Top right-hand panel: the signal-to-noise ratio of the inverted
source intensity.
Bottom left-hand panel: the simulated data.  Bottom middle panel: the reconstructed image
using the most likely reconstructed source (top left-hand panel) and the
$\responseSet$ matrix from the potential and PSF models.  Reconstructed data
is confined to an annular region that maps on to the source plane.  Bottom right-hand panel: the
residual image obtained by subtracting the bottom middle panel from the bottom left-hand panel. 
The interior of the annular region is less noisy than the exterior, indicating
that the reconstructed image is fitting to the noise in the simulated data.}
\end{figure*}
\begin{figure*}
\vspace{0.0in}
\includegraphics[width=170mm]{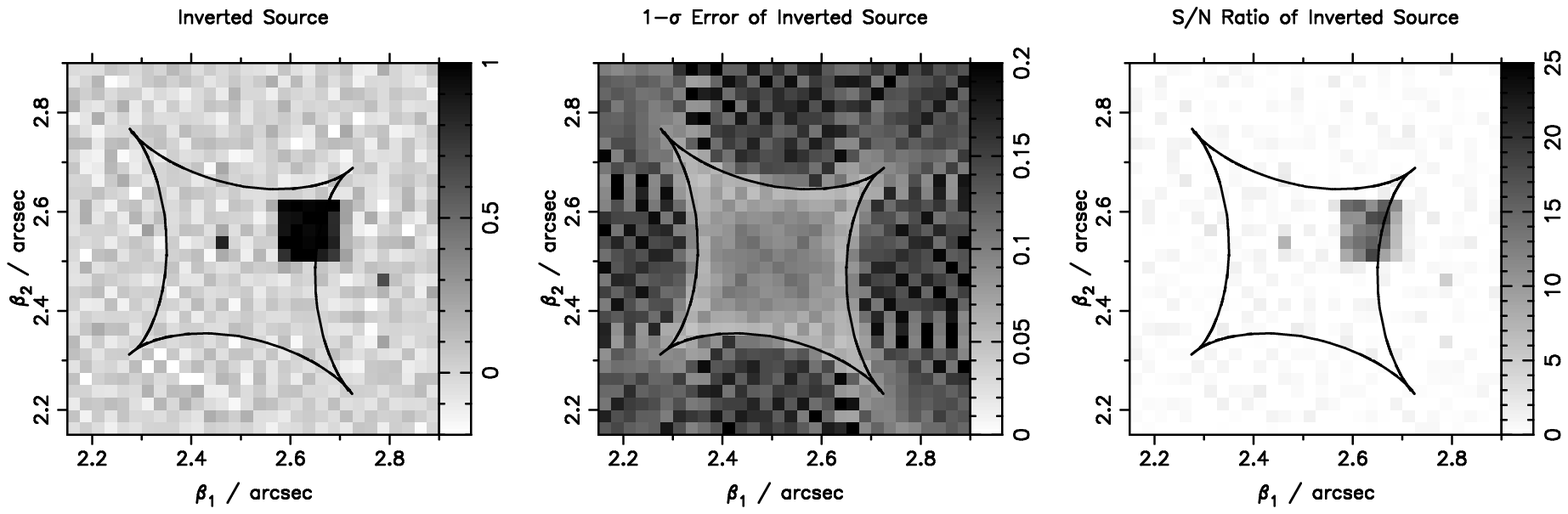}
\vspace{0.1in}
\includegraphics[width=170mm]{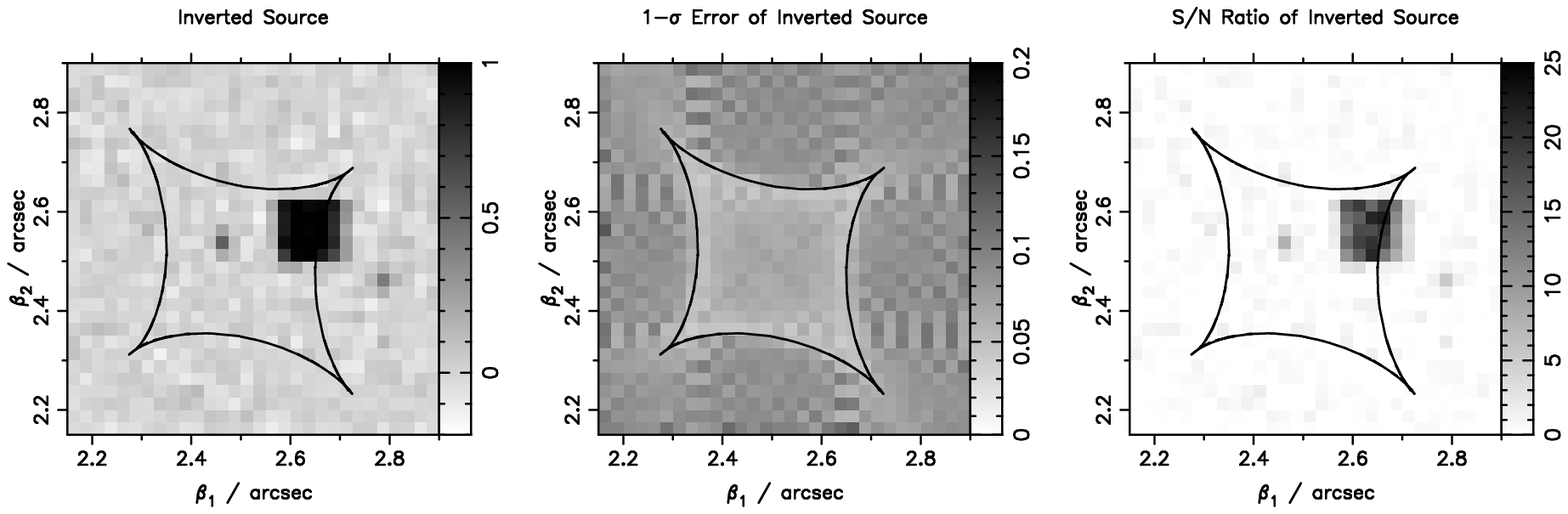}
\vspace{0.1in}
\includegraphics[width=170mm]{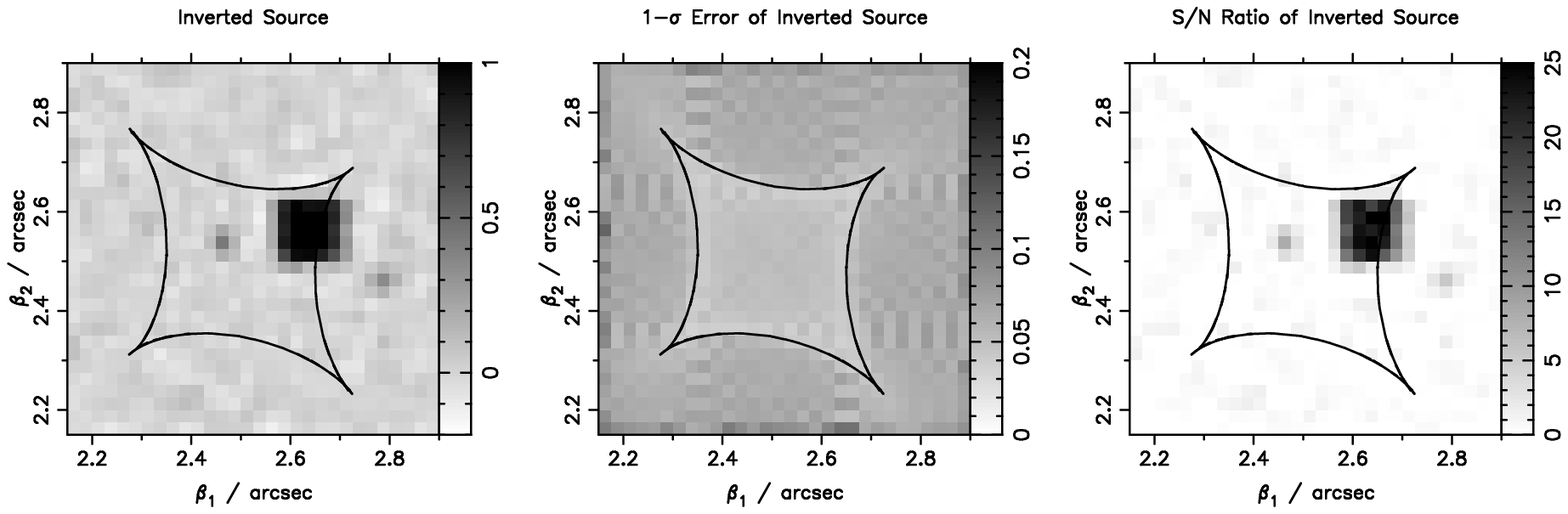}

\caption{\label{fig:RSr17}  The regularised source inversions of box and point
sources with zeroth-order, gradient and curvature regularisations.  Top row, from
left to right: most probable inverted source, the 1-$\sigma$ error, and the
signal-to-noise ratio with zeroth-order regularisation.  Middle row, from left to
right: same as top row but with gradient regularisation.  Bottom row, from
left to right: same as top row but with curvature regularisation.  The panels
in each column are plotted on the same scales for comparison among the
different forms of regularisation.}

\end{figure*}
\begin{figure*}
\vspace{0.0in}
\includegraphics[width=170mm]{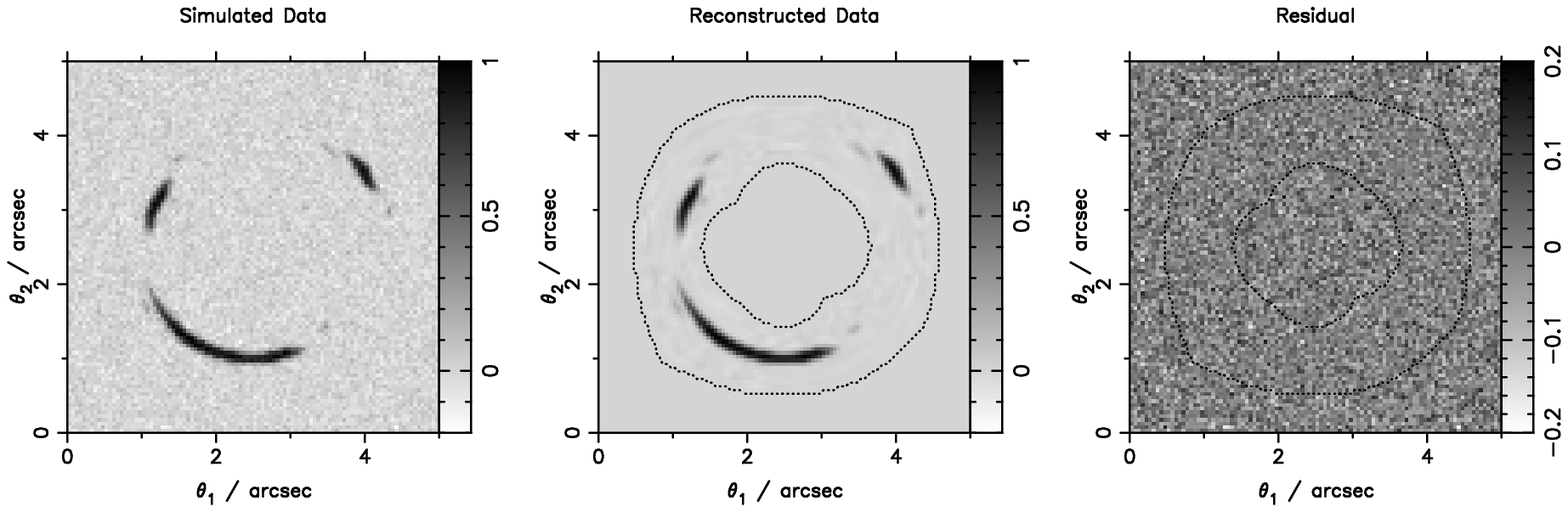}

\caption{\label{fig:RRes17} The image residual for gradient regularised source
inversion with box and point sources.  From left to right: simulated data,
reconstructed data using the corresponding most probable inverted source in
Fig.~\ref{fig:RSr17}, and the residual equalling the difference between
simulated and reconstructed data.  The reconstructed data is restricted to the
annulus marked by dotted lines that is mapped from the finite source grid
using $\responseSet$.  The noise in the residual image is more uniform
compared to that of the unregularised inversion in Fig.~\ref{fig:NRSr17}.}

\end{figure*}

For regularised inversions, we solve equation (\ref{eq:OptLam}) for
the optimal regularisation constant for each of the three forms of
regularisation.  We list the optimal regularisation constants, $\hat\lambda$,
and the associated log evidence evaluated at $\hat\lambda$ in Table
\ref{tab:OptLambda17}.  Fig.~\ref{fig:RSr17} shows the most probable inverted
source using the optimal regularisation constant in Table
\ref{tab:OptLambda17} for each of the three forms of regularisation.  By
visual inspection, the inverted source intensities (left-hand panels) with gradient
regularisation matches the original source brightness distribution
(Fig.~\ref{fig:simData17}) the best since curvature regularisation
overly-smears the sharp edges and zeroth-order regularisation leads to higher
background noise.  This is supported quantitatively by the values of the evidence
in Table \ref{tab:OptLambda17} with the highest value for gradient
regularisation (which is $\sim \mathrm{e}^{37}$ more probable than curvature
regularisation and $\sim \mathrm{e}^{222}$ more probable than zeroth-order
regularisation).  Again, this example illustrates that the
signal-to-noise ratio does not determine the optimal regularisation - the
right-hand panels of Fig.~\ref{fig:RSr17} show that curvature regularisation leads
to the highest signal-to-noise ratio, but the Bayesian analysis objectively
ranks gradient over curvature!  Finally, Fig.~\ref{fig:RRes17} shows the
reconstructed image (middle panel) and the image residual (right-hand panel) using
the gradient regularisation.  The corresponding plots for the zeroth-order and
curvature regularisations are similar and hence are not shown.

\begin{table}
\caption{\label{tab:OptLambda17} The optimal regularisation constant for each
of the three forms of regularisation for the inversion of box and point
sources.  The log evidence evaluated at the optimal regularisation constant is
also listed.}

\begin{tabular}{|c|c|c|c|}
\hline
Regularisation  & zeroth-order & gradient & curvature \\
\hline 
$\hat{\lambda}$ & 19.8 & 21.0 & 17.1 \\
$\log P(\dataVec|\hat{\lambda}, \responseSet, \regSet)$ & 6298 & 6520 & 6483 \\
\hline
\end{tabular}
\end{table}

\subsection{Discussion}
\label{sect:appgl:discussion}

\subsubsection{Preferred form of regularisation}
\label{sect:appgl:discussion:prefReg}

The two examples of source inversion considered in Sections \ref{sect:appgl:demo1} and \ref{sect:appgl:demo2} show that the
form of regularisation that is optimally selected in the Bayesian approach
depends on the nature of the source.  Generally, with the three forms of
regularisation considered, curvature regularisation is preferred for smooth
sources and gradient (or even zeroth-order) is preferred for sources with sharp
intensity variations.  In the two examples of source inversion, we found that
at least one of the three considered forms of regularisation (which is not
always the curvature form) allowed us to reconstruct successfully the original
source in the inversion.  Therefore, we did not need to consider other forms
of regularisation.  Nonetheless, this does not preclude other forms of
regularisation to be used.  Even with additional types of regularisation,
Bayesian analysis can always be used to choose the optimal one from the
selection of forms of regularisation.


\subsubsection{Optimal number of source pixels}
\label{sect:appgl:discussion:OptNs}

So far, we have not discussed the size and the region of the source pixels to
use.  In both demonstration examples in Sections \ref{sect:appgl:demo1} and \ref{sect:appgl:demo2}, we used source
pixels that were half the size of the image pixels.  In reality, one has to
find the source region and the size of source pixels to use.  

The selection of the source pixel size for a given source region can be
accomplished using Bayesian analysis in the model comparison step of Section
\ref{sect:BayesInf:ModelComp:rankModel} (the size of the source pixels is part of $\responseSet$ since different
source pixels sizes result in different matrices $\responseSet$).  We find
that source pixels sizes that are too large do not have enough degrees of
freedom to fit to the data.  On the other hand, source pixels that are too
small will result in some source pixels being excluded in the $\responseSet$
matrix (using the $\responseSet$ construction method in \citet{TK04}), which
leads to a failure in the most likely source inversion since some pixels will
be unconstrained.  Therefore, for fixed pixel sizes over a source region
(which our codes assume), the minimum source pixel size will be set by the
minimum magnification over the source region. To improve the resolution in
areas where there is more information, one would need to use adaptive grids. 
\citet{DW05} have used adaptive grids in their source inversion routine, and
we are also in the process of developing a code with adaptive gridding that
will appear in a future paper.  Our methods differ from that of \citet{DW05}
in that we follow a Bayesian approach and can thus quantitatively compare the
forms of regularisation and the structure of source pixellation. 

At this stage, we cannot compare different source regions since the annular
region on the image plane that maps to the source plane changes when the
source region is altered.  Recall that we only use the data within the annulus
for source inversion.  If the annular region changes, the data for inversion
also changes.  For model comparison between different data sets, we would need
to know the normalisation in equation (\ref{eq:postAR}), which we do not. 
Therefore, the best we can do in terms of source region selection is to pick a
region that is large enough to enclose the entire luminous source, but small
enough to not have the corresponding annular region exceeding the image region
where we have data.  Once the source region is selected, we can apply Bayesian
analysis to determine the optimal source pixel size (subject to the minimum
limit discussed above) and the optimal form of regularisation given the data.


\section{Conclusions and further work}
\label{sect:conclusion}

We introduced and applied Bayesian analysis to the problem of regularised
source inversion in strong gravitational lensing.  In the first level of
Bayesian inference, we obtained the most probable inverted source of a given
lens potential and PSF model $\responseSet$, a given form of regularisation
$\regSet$ and an associated regularisation constant $\lambda$; in the second
level of inference, we used the evidence $P(\dataVec|\lambda,
\responseSet,\regSet)$ to obtain the optimal $\lambda$ and rank the different
forms of regularisation, assuming flat priors in $\lambda$ and $\regSet$.

We considered three different types of regularisation (zeroth-order, gradient and
curvature) for source inversions.  Of these three, the preferred form of regularisation depended
on the intrinsic shape of the source intensity distribution: in general, the
smoother the source, the higher the derivatives of the source intensity in the
preferred form of regularisation.  In the demonstrated examples of first two Gaussian
sources, and then a box with point sources, we optimised the evidence
$P(\dataVec|\lambda, \responseSet,\regSet)$ and numerically solved for the
regularisation constant for each of the three forms of regularisation.  By
comparing the evidence of each regularisation
evaluated at the optimal $\lambda$, we found that the curvature regularisation
was preferred with the highest value of evidence for the two Gaussian sources,
and gradient regularisation was preferred for the box with point sources.

The study of the three forms of regularisation demonstrated the Bayesian
technique used to compare different regularisation schemes objectively.  The
method is general, and the evidence can be used to rank other forms of
regularisation, including non-quadratic forms (e.g. maximum entropy methods) that lead to non-linear inversions
\citep[e.g.\ ][]{W96, WWLH05, BL05}.  We restricted ourselves to linear inversion
problems with quadratic forms of regularising function for computational
efficiency.  

In the demonstration of the Bayesian technique for regularised source
inversion, we assumed Gaussian noise, which may not be applicable to real
data.  In particular, Poisson noise may be more appropriate for real data, but the use of
Poisson noise distributions would lead to non-linear inversions that we tried
to avoid for computational efficiency.  Nonetheless, the Bayesian method of
using the evidence to rank the different models (including noise models) is
still valid, irrespective of the linearity in the inversions.

We could also use Bayesian analysis to determine the optimal size of source
pixels for the reconstruction.  The caveat is to ensure that the annular
region on the image plane where the source plane maps is unchanged for
different pixel sizes.  Currently the smallest pixel size is limited by the
region of low magnifications on the source plane.  In order to use smaller
pixels in regions of high magnifications, adaptive source gridding is needed. 
This has been studied by \citet{DW05}, and we are currently upgrading our
codes to include this.

The Bayesian approach can also be applied to potential reconstruction on a
pixellated potential grid.  \citet*{BSK01} proposed a method to perturbatively
and iteratively correct the lens potential from a starting model by solving a
first order partial differential equation.  This method has been studied by
\citet{K05} and \citet{SB05}.  The perturbation differential equation can be
written in terms of matrices for a pixellated source brightness distribution and a
pixellated potential, and the potential correction of each iteration can be
obtained via a linear matrix inversion.  This pixellated potential
reconstruction is very similar to the source inversion problem and we are
currently studying it in the Bayesian framework.

The Bayesian analysis introduced in this paper is general and was so naturally
applicable to both the source and potential reconstructions in strong
gravitational lensing that we feel the Bayesian approach could be useful in
other problems involving model comparison.


\section*{Acknowledgments}

We thank D. MacKay and S. Warren for useful discussions and encouragement, and
the referee L. Koopmans for both his guidance on the methodology, and his very
constructive comments that greatly improved the presentation of this work.
This work was supported by the NSF under award AST05-07732 and in part by the
U.S. Department of Energy under contract number DE-AC02-76SF00515.  SS
acknowledges the support of the NSERC (Canada) through the Postgraduate
Scholarship.


\bibliography{BayesSr}
\bibliographystyle{mn2e}


\appendix


\section[]{Forms of regularisation}
\label{app:regForms}

We consider the three most common quadratic functional forms of the
regularisation found in the local literature: 
``zeroth-order,'' ``gradient,'' and ``curvature'' \citep[][\S18.4 and \S18.5]{P92}.
For clarity reasons, we use explicit index and summation notation instead of vector and matrix notation for the expression of the regularising function $E_\mathrm{S}(\mathbfit{s})$.

Zeroth-Order regularisation is the simplest case.  The functional form is
\be
\label{eq:ESquad}
E_{\mathrm{S}}(\mathbfit{s}) = \frac{1}{2} \sum_{i=1}^{N_{\rm s}}  s_i^2,
\ee
and its Hessian is the identity operator $\mathbfss{C}=\mathbfss{I}$.  This
form of regularisation tries to minimize the intensity at every source pixel
as a way to smooth the source intensity distribution. It introduces no
correlation between the reconstruction pixel values.

To discuss gradient and curvature forms of regularisation, we label the
pixels by their x and y locations (i.e., have two labels $(i_1, i_2)$ for each
pixel location instead of only one label $(i)$ as in Section \ref{sect:appgl:reg}) since the
mathematical structure and nomenclature of the two forms of regularisation are
clearer with the two-dimensional labelling.  Let $s_{i_1,i_2}$ be the source
intensity at pixel $(i_1,i_2)$, where $i_1$ and $i_2$ range from
$i_1=1,\ldots,N_{\rm 1s}$ and $i_2=1,\ldots,N_{\rm 2s}$.  The total number of source pixels is thus
$N_{\rm s}=N_{\rm 1s} N_{\rm 2s}$.  It is not difficult to translate the labelling of pixels
on a rectangular grid from two dimensions to one dimension for Bayesian
analysis.  For example, one way is to let $i=i_1+(i_2-1) N_{\rm 2s}$. 

A form of gradient regularisation is 
\bea
\label{eq:ESgrad}
E_{\mathrm{S}}(\mathbfit{s}) &=& \phantom{+} \frac{1}{2} \sum_{i_1=1}^{N_{\rm 1s}-1} \sum_{i_2=1}^{N_{\rm 2s}} \left[s_{i_1,i_2}-s_{i_1+1,i_2}\right]^2 
\nonumber \\ & &
+ \frac{1}{2} \sum_{i_1=1}^{N_{\rm 1s}} \sum_{i_2=1}^{N_{\rm 2s}-1} \left[s_{i_1,i_2}-s_{i_1,i_2+1}\right]^2 
\nonumber \\ & &
+ \frac{1}{2} \sum_{i_1=1}^{N_{\rm 1s}} s_{i_1,N_{\rm 2s}}^2 + \frac{1}{2} \sum_{i_2=1}^{N_{\rm 2s}} s_{N_{\rm 1s},i_2}^2.
\eea
The first two terms are proportional to the gradient values of the pixels, so
this form of regularisation tries to minimize the difference in the intensity
between adjacent pixels.  The last two terms can be viewed as gradient terms if we
assume that the source intensities outside the grid are zeros.  Although the
non-singularity of the Hessian of $E_{\mathrm{S}}$ is not required for equation (\ref{eq:sMP}) since equation
(\ref{eq:ESgrad}) is of the form $E_{\mathrm{S}}(\mathbfit{s}) = \frac{1}{2}
\mathbfit{s}^{\mathrm{T}} \mathbfss{C} \mathbfit{s}$, these last two terms
ensure that the Hessian of $E_{\mathrm{S}}$ is non-singular and lead to
$\mathbfit{s}_{\mathrm{reg}}=\mathbf{0}$.  The non-singularity of the Hessian of $E_{\mathrm{S}}$  (i.e., $\rm{det} \mathbfss{C} \neq 0$) is crucial to the model comparison process described in Section \ref{sect:BayesInf:ModelComp:rankModel} that requires the evaluation of the log evidence in equation (\ref{eq:EvidFull}).

A form of curvature regularisation is
\bea
\label{eq:EScurv}
E_{\mathrm{S}}(\mathbfit{s}) &=& \phantom{+} \frac{1}{2} \sum_{i_1=1}^{N_{\rm 1s}-2} \sum_{i_2=1}^{N_{\rm 2s}} \left[s_{i_1,i_2}-2s_{i_1+1,i_2}+s_{i_1+2,i_2}\right]^2  
 \nonumber \\ & & 
+ \frac{1}{2} \sum_{i_1=1}^{N_{\rm 1s}} \sum_{i_2=1}^{N_{\rm 2s}-2} \left[s_{i_1,i_2}-2s_{i_1,i_2+1}+s_{i_1,i_2+2}\right]^2  
 \nonumber \\ & & 
+ \frac{1}{2} \sum_{i_1=1}^{N_{\rm 1s}} \left[s_{i_1,N_{\rm 2s}-1} - s_{i_1,N_{\rm 2s}}\right]^2  \nonumber \\ & &
+ \frac{1}{2} \sum_{i_2=1}^{N_{\rm 2s}} \left[s_{N_{\rm 1s}-1,i_2} - s_{N_{\rm 1s},i_2}\right]^2  \nonumber \\ & & 
+ \frac{1}{2} \sum_{i_1=1}^{N_{\rm 1s}} s_{i_1,N_{\rm 2s}}^2 + \frac{1}{2} \sum_{i_2=1}^{N_{\rm 2s}} s_{N_{\rm 1s},i_2}^2.
\eea
The first two terms measure the second derivatives (curvature) in the x and y
directions of the pixels.  The remaining terms are added to enforce our a
 priori preference towards a blank image with non-singular Hessian (important for the model ranking) that gives
 $\mathbfit{s}_{\mathrm{reg}}=\mathbf{0}$.  In essence, the majority of the source pixels have
curvature regularisation, but two sides of the bordering pixels that do not
have neighbouring pixels for the construction of curvature terms have gradient
and zeroth-order terms instead.

It is not difficult to verify that all three forms of regularisation have
$\mathbfit{s}_{\mathrm{reg}}=\mathbf{0}$ in the expansion in equation
(\ref{eq:EStaylor}).  
Therefore, equation
(\ref{eq:sMP}) for the most probable solution is applicable, as asserted in
Section \ref{sect:appgl:reg}. 

None of the three forms of regularisation impose the source intensity to be
positive.  In fact, equations (\ref{eq:ESquad}) to (\ref{eq:EScurv}) suggest
that the source intensities are equally likely to be positive or negative
based on only the prior. 

In principle, one can continue the process and construct regularisations of
higher derivatives.  Regularisations with higher derivatives usually imply
smoother source reconstructions, as the correlations introduced by the
gradient operator extend over larger distances. 
Depending on the nature of the source,
regularisations of higher derivatives may not necessarily be preferred over those
of lower derivatives: astronomical sources tend to be fairly compact.  
Therefore, we restrict ourselves to the  three lowest
derivative forms of the regularisation for the source inversion problem.


\section[]{Explanation of the source covariance matrix in Bayesian
  analysis}
\label{app:SrCovM}

\subsection*{Notation}
Expressed in terms of matrix and vector multiplications, recall equation
(\ref{eq:dj}) for the image intensity vector is 
\be
\label{eq:App:data}
\dataVec=\responseSet \srVec + \noiseVec,
\ee
where $\responseSet$ is the lensing (response) matrix, $\srVec$ is the source intensity
vector and $\noiseVec$ is the noise vector.  Recall equation (\ref{eq:ED})
is
\be
\label{eq:App:ED}
E_{\mathrm{D}}(\srVec) = \frac{1}{2}(\responseSet \srVec-\dataVec)^{\rm T}
\imCM^{-1}(\responseSet \srVec - \dataVec), 
\ee
where $\imCM=\langle \noiseVec \noiseVec^{\rm T} \rangle$ is the image noise
covariance matrix.  We write the prior exponent as 
\be
\label{eq:App:ES}
\lambda E_{\mathrm{S}}(\srVec) = \frac{1}{2}\srVec^{\rm T}\priorCM^{-1}\srVec,
\ee
where, for simplicity, we have set $\srVec_{\rm reg}=\mathbf{0}$ and
$E_{\mathrm{S}}(\mathbf{0})=0$ (valid for the regularisation schemes considered
  in Appendix \ref{app:regForms}), and $\priorCM =\langle \srVec
  \srVec^{\rm T}\rangle$ is the a priori source covariance  
matrix.  Comparing to equation (\ref{eq:EStaylor}), $\priorCM=(\lambda
\hessS)^{-1}$.  Combining equations (\ref{eq:App:ED}) and (\ref{eq:App:ES}),
the exponent of the posterior is 
\bea
\label{eq:App:M}
M(\srVec) &=& E_{\mathrm{D}}(\srVec)+\lambda E_{\mathrm{S}}(\srVec) 
\nonumber \\ &=& \frac{1}{2}(\responseSet \srVec-\dataVec)^{\rm T}
\imCM^{-1}(\responseSet \srVec-\dataVec) + \frac{1}{2}\srVec^{\rm
  T}\priorCM^{-1}\srVec. 
\eea

\subsection*{Most likely estimate}
The most likely estimate is given by $\nabla E_{\mathrm{D}}(\srMLVec)=\mathbf{0}$, which gives
\be
\label{eq:App:gradgradED}
\responseSet^{\rm T} \imCM^{-1} (\responseSet \srMLVec - \dataVec)=\mathbf{0}.
\ee
Rearranging the previous equation, we obtain
\be
\label{eq:App:sMLinit}
\srMLVec = (\responseSet^{\rm T} \imCM^{-1}
\responseSet)^{-1}\responseSet^{\rm T} \imCM^{-1} \dataVec.
\ee
Differentiating $E_{\mathrm{D}}(\srVec)$ again gives the Hessian
\be
\label{eq:App:hessB}
\hessD\equiv \nabla\nabla E_{\mathrm{D}}(\srVec) = \responseSet^{\rm T} \imCM^{-1} \responseSet. 
\ee
This in turn allows us to write
\be
\label{eq:App:sML}
\srMLVec = \hessD^{-1}\responseSet^{\rm T} \imCM^{-1} \dataVec,
\ee
which is equation (\ref{eq:sML}).

By construction, $\imCM$, $\priorCM$, and $\hessD$ are symmetric matrices.

\subsection*{Error on most likely estimate}
Let us assume that the true source intensity is $\srVec_*$ (i.e. the actual
true source intensity for the particular image we are considering). Now
consider the expectation value of $\srMLVec$ over realisations of the noise
$\noiseVec$: 
\be
\label{eq:App:sMLavg}
\langle \srMLVec \rangle 
= \hessD^{-1}\responseSet^{\rm T} \imCM^{-1}\langle \responseSet \srVec_* +
\noiseVec \rangle
= \hessD^{-1}\responseSet^{\rm T} \imCM^{-1}\responseSet \srVec_* = \srVec_*,
\ee
where we have used $\langle \noiseVec \rangle = \mathbf{0}$ and angle brackets denote
averages over noise realisations. Thus, we see that $\srMLVec$ is an
{\em unbiassed} estimator of $\srVec_*$.

Now consider the covariance of $\srMLVec$. Since 
$\langle \srMLVec \rangle = \srVec_*$, the covariance is given by 
\bea
\label{eq:App:sMLcov}
\langle (\srMLVec-\srVec_*)(\srMLVec-\srVec_*)^{\rm T}\rangle
& = &  \langle \srMLVec\srMLVec^{\rm T}\rangle + \srVec_*\srVec_*^{\rm T}
\nonumber \\ 
& & -\srVec_*\langle \srMLVec^{\rm T}\rangle - \langle \srMLVec\rangle
\srVec_*^{\rm T} \nonumber \\
& = & \langle \srMLVec\srMLVec^{\rm T}\rangle - \trueCM.
\eea
where $\trueCM=\srVec_*\srVec_*^{\rm T}$ is the covariance matrix of the true
signal and, once again, angle brackets denote averages over noise
realisations. The term $\langle \srMLVec\srMLVec^{\rm T}\rangle$
above is given by
\bea 
\label{eq:App:sMLcovTerm1}
\langle \srMLVec\srMLVec^{\rm T}\rangle & = &
\hessD^{-1}\responseSet^{\rm T}\imCM^{-1}\langle \dataVec \dataVec^{\rm
T}\rangle \imCM^{-1}\responseSet \hessD^{-1} 
\nonumber \\ & = & \hessD^{-1}\responseSet^{\rm T} \imCM^{-1}\langle
(\responseSet \srVec_*+\noiseVec)(\responseSet \srVec_*+\noiseVec)^{\rm
  T}\rangle \imCM^{-1}\responseSet \hessD^{-1} 
\nonumber \\ & = & \hessD^{-1}\responseSet^{\rm T} \imCM^{-1} (\responseSet
\srVec_*\srVec_*^{\rm T}\responseSet^{\rm T}+\imCM) \imCM^{-1}\responseSet
\hessD^{-1} 
\nonumber \\ & = & \hessD^{-1}\hessD\trueCM \hessD\hessD^{-1} +
\hessD^{-1}\hessD\hessD^{-1} 
\nonumber \\ & = & \trueCM + \hessD^{-1}.  
\eea
Inserting equation (\ref{eq:App:sMLcovTerm1}) in (\ref{eq:App:sMLcov}), the
covariance of $\srMLVec$ is given simply by 
\be
\label{eq:sMLcovResult}
\langle (\srMLVec-\srVec_*)(\srMLVec-\srVec_*)^{\rm T}\rangle = \hessD^{-1},
\ee
which agrees with equation (\ref{eq:sCov}) since $\hessM=\hessD$ for the
most likely solution (with $\lambda=0$).

\subsection*{Most probable estimate}

The most probable estimate is given by $\nabla M(\srMPVec) =\mathbf{0}$,
which gives 
\be
\label{eq:App:gradgradM}
\responseSet^{\rm T} \imCM^{-1} (\responseSet
\srMPVec-\dataVec)+\priorCM^{-1}\srMPVec=\mathbf{0}.
\ee
Rearranging, we get
\be
\label{eq:App:sMPinit}
\srMPVec = (\priorCM^{-1}+\responseSet^{\rm T}
\imCM^{-1}\responseSet)^{-1}\responseSet^{\rm T} \imCM^{-1}\dataVec.
\ee
Differentiating $M(\srVec)$ again gives the Hessian
\be
\label{eq:App:hessM}
\hessM\equiv \nabla\nabla M(\srVec) = \priorCM^{-1}+\responseSet^{\rm T} \imCM^{-1}\responseSet = \priorCM^{-1}+\hessD, 
\ee
which, in turn, allows us to write
\be
\label{eq:App:sMP}
\srMPVec = \hessM^{-1}\responseSet^{\rm T} \imCM^{-1}\dataVec 
= \hessM^{-1}\hessD\hessD^{-1}\responseSet^{\rm T} \imCM^{-1}\dataVec = \hessM^{-1}\hessD\srMLVec,
\ee
which agrees with equation (\ref{eq:sMP}).

The Hessian A is symmetric by construction.

\subsection*{Error on MP estimate}

Let us again assume that the true source intensity is $\srVec_*$. Using equations
(\ref{eq:App:sMP}) and (\ref{eq:App:sMLavg}), the expectation value of
$\srMPVec$ over realisations of the noise $\noiseVec$ is 
\be
\label{eq:App:sMPavg}
\langle \srMPVec \rangle = \hessM^{-1}\hessD \langle \srMLVec\rangle = \hessM^{-1}\hessD\srVec_*,
\ee
where angle brackets denote averages over noise realisations. Thus, we
see that $\srMPVec$ is a {\em biassed} estimator (in general) of
$\srVec_*$. We must therefore be careful when considering errors.

First consider the covariance of $\srMPVec$, which is given by
\bea
\label{eq:App:sMPcovWD} 
\langle (\srMPVec-\langle \srMPVec\rangle)
(\srMPVec-\langle \srMPVec\rangle)^{\rm T}\rangle  
& = & \hessM^{-1}\hessD\hessM^{-1}, 
\eea
where we have used equations (\ref{eq:App:sMP}), (\ref{eq:App:sMPavg}) and (\ref{eq:App:sMLcovTerm1}).
Remembering that $\hessM=\priorCM^{-1}+\hessD$, we have $\hessD=\hessM-\priorCM^{-1}$, so the final
result is
\be
\label{eq:App:sMPcovWDresult}
\langle (\srMPVec-\langle \srMPVec\rangle)
(\srMPVec-\langle \srMPVec\rangle)^{\rm T}\rangle = 
\hessM^{-1}-\hessM^{-1}\priorCM^{-1}\hessM^{-1},
\ee
which is equivalent to the equation (17) in \citet{WD03}.

We verified equation (\ref{eq:App:sMPcovWDresult}) by a Monte Carlo simulation
of 1000 noise realisations of the source brightness distribution described in Section
\ref{sect:appgl:demo1:simData}.  The noise realisations differ only in the values of the random seed
used to generate random noise in the simulated data.  We used curvature
regularisation (see Appendix \ref{app:regForms}) with a fixed (and nearly optimal) value of
the regularisation constant $\lambda$ for each of the 1000 source inversions.
The standard deviation of $\srMPVec$ calculated from the 1000 inverted source
distributions agrees with the 1-$\sigma$ error from equation
(\ref{eq:App:sMPcovWDresult}).

Equation (\ref{eq:App:sMPcovWDresult}) gives the error from the {\em
reconstructed source} $\srMPVec$.  
Since $\srMPVec$ is a biassed estimator of $\srVec_*$,
what we really want to know is not the covariance above, but the
quantity $\langle (\srMPVec- \srVec_*)(\srMPVec-\srVec_*)^{\rm T}\rangle$,
which gives us the distribution of errors from the {\em true
source}. This is given by
\bea
\label{eq:App:sMPcov}
\langle (\srMPVec- \srVec_*)(\srMPVec-\srVec_*)^{\rm T}\rangle
& = & \hessM^{-1}\hessD \trueCM \hessD \hessM^{-1} +\hessM^{-1}\hessD
\hessM^{-1} \nonumber \\
& & + \trueCM - \trueCM \hessD\hessM^{-1} \nonumber \\
& & - \hessM^{-1}\hessD\trueCM,
\eea
where we have again used equations (\ref{eq:App:sMP}), (\ref{eq:App:sMPavg}) and (\ref{eq:App:sMLcovTerm1}).
Substituting $\hessD=\hessM-\priorCM^{-1}$ gives, after simplifying, 
\bea
\label{eq:App:sMPcovResult}
\langle (\srMPVec- \srVec_*)(\srMPVec-\srVec_*)^{\rm T}\rangle
& = & \hessM^{-1} + \hessM^{-1}\priorCM^{-1}\nonumber \\
& & (\trueCM \priorCM^{-1}-\IdenM)\hessM^{-1}.
\eea
In reality, we do not know $\trueCM$ (as this would require
knowing the true source intensity $\srVec_*$). However, by averaging over
source brightness distributions (denoted
by a bar), we have $\overline{\trueCM}=\priorCM$. 
This is the manifestation of our explicit assumption that all source intensity
distributions are drawn from the prior probability density defined by
equation~(\ref{eq:prior}). 
Thus,
\be
\label{eq:App:sMPcovAvgIm}
\overline{\langle (\srMPVec- \srVec_*)(\srMPVec-\srVec_*)^{\rm T}\rangle}=\hessM^{-1},
\ee
which is the inverse of $\nabla\nabla M(\srVec)$.  In words, the
covariance matrix describing the uncertainties in the inverted source intensity is given by
the width of the approximated Gaussian posterior in equation
(\ref{eq:posterior2}), which is $\mathbfss{A}^{-1}$.  The covariance matrix of
$\srMPVec$ in equation (\ref{eq:App:sMPcovWDresult}) in general
under-estimates the error relative to the true source image because it does not incorporate the bias in the
reconstructed source.


\bsp
\label{lastpage}
\end{document}